\documentstyle[11pt,aaspp4]{article}
\def\msun{M_{\odot}}
\def\Lsun{L_{\odot}}
\def\rsun{R_{\odot}}
\def\gapprox{\;\rlap{\lower 3.0pt                       
             \hbox{$\sim$}}\raise 2.5pt\hbox{$>$}\;}
\def\lapprox{\;\rlap{\lower 3.0pt                       
             \hbox{$\sim$}}\raise 2.5pt\hbox{$<$}\;}
\newcommand{\br}{{\bf r}}
\newcommand{\brd}{\dot{\bf r}}
\begin{document}
\title{The Effect of Tidal Inflation Instability on the Mass and Dynamical 
Evolution of Extrasolar Planets with Ultra-Short Periods}
\author{Pin-Gao Gu}
\affil{Institute of Astronomy and Astrophysics, Academia Sinica,
Taipei 106, Taiwan, R.O.C.}
\and
\author{Douglas N. C. Lin and Peter H. Bodenheimer}
\affil{UCO/Lick Observatory, University of California, Santa Cruz,
CA 95064, U.S.A.}
\centerline{\today}

\begin{abstract}
We investigate the possibility of substantial inflation of
short-period Jupiter-mass planets, as a result of their internal tidal
dissipation associated with the synchronization and circularization of
their orbits. 
We employ the simplest 
prescription based on an equilibrium model with a constant lag
angle for all components of the tide.
We show that 1) in the low-eccentricity limit, the
synchronization of the planets' spin with their mean motion is
established before tidal dissipation can significantly  modify their
internal structure.  2) But, above a critical eccentricity, which is a
function of the planets' semimajor axis, tidal dissipation of energy
during the circularization process can induce planets to inflate in
size before their eccentricity is damped.  3) For moderate
eccentricities, the planets adjust to stable thermal equilibria in
which the rate of their internal tidal dissipation is balanced by the
enhanced radiative flux associated with their enlarged radii.  4) For
sufficiently large eccentricities, the planets swell beyond two
Jupiter radii and their internal degeneracy is partially lifted.
Thereafter, their thermal equilibria become unstable and they undergo
runaway inflation until their radii exceed the Roche radius.  5) We
determine the necessary and sufficient condition for this tidal
inflation instability.  6) These results are applied to study
short-period planets.  We show that for young Jupiter-mass planets,
with a period less than 3 days, an initial radius about 2 Jupiter
radii, and an orbital eccentricity greater than 0.2, the energy
dissipated during the circularization of their orbits is sufficiently
intense and protracted to inflate their sizes up to their Roche radii.
7) We estimate the mass loss rate, the asymptotic planetary masses,
and the semi-major axes for various planetary initial orbital
parameters.  The possibility of gas overflow through both inner (L1)
and outer (L2) Lagrangian points for the planets with short periods or
large eccentricities is discussed.  8) Planets with more modest
eccentricity ($< 0.3$) and semi-major axis ($>0.03-0.04$ AU) lose mass
via Roche-lobe overflow mostly through the inner Lagrangian (L1)
point.  Due to the conservation of total angular momentum, these
mass-losing planets migrate outwards, such that their semi-major axes
and Roche radii increase while their mass, eccentricity, and tidal
dissipation rate decrease until the mass loss is quenched.  9) Based
on these results, we suggest that the combined effects of
self-regulated mass loss and tidally driven orbital evolution may be
responsible for the apparent lack of giant planets with ultra-short
periods $\lesssim 3$ days.  10) Mass loss during their orbital
circularization may also have caused the planets with periods in the
range $\sim 3-7$ days to be less massive than long-period planets
which are not affected by the star-planet tidal interaction.  11) The
accretion of the short-period planets' tidal debris can also lead to
the surface-layer contamination and metallicity enhancement of their
host stars.  12) Among the planets with periods of 1-3 weeks today,
some may have migrated outwards and attained circular orbits while
others may have preserved their initial eccentricity and semimajor
axis.  Therefore, planets with circular orbits are expected to coexist
with those with eccentric orbits in this period range.  13) Gross
tidal inflation of planets occurs  on the time scale $\sim 10^{6}$ yrs
after their formation for a brief interval of $\sim 10^5$ yrs.  The
relatively large sizes of their classical and weak-line T Tauri host
stars increases the planets' transit probability.  The inflated sizes
of the tidally heated planets also increases the eclipse depth of such
transit events.  Thus, the tidal inflation and disruption of  planets may
be directly observable around classical and weak-line T Tauri stars.

\end{abstract}

\section{Introduction}
Recent searches for planets around nearby solar type stars, based on
the radial velocity method, show a roughly logarithmic period
distribution in the currently observed period range of 3 days to
several years (\cite{marcy00}). The absence of ultra-short-period
Jupiter-mass planets with $P<3$ days, which corresponds to an orbital
semi-major axis $a$ of 0.04 AU, appears to be real and not an
observational selection effect (Cumming {\it et al.} 1999).  The host
stars of any ultra-short-period planets are expected to have large
radial velocity modulations which would make them conspicuous.
Close-in planets also have greater transit probabilities.  The
unsuccessful photometric search for transiting planets around main
sequence stars in the core of the globular cluster 47 Tuc
(\cite{47Tuc}) provides more evidence for a lack of Jovian planets
with ultra-short orbital periods.  Thus, the cutoff in the extrasolar
planets' period distribution is likely to be caused by some physical
effect.  For example, it has been suggested (\cite{kl02}) that the
inner disk might be hot enough to allow the magneto-rotational
instability to operate and thereby is largely depleted, terminating
the inward migration through planet-disk interaction and causing a
lack of giant planets with orbital periods less than 3 days. In the
case of 47 Tuc,  most of the  Jovian planets might
have been expelled from their host extrasolar systems due to frequent
stellar encounters in the dense stellar environment (\cite{ds01}), or
they may have had difficulties in  forming  in circumstellar disks with low
metallicities (\cite{pol96}).

In this contribution, we examine the effect of tidal interaction
between ultra-short-period planets and their host stars and explore
the possibility that this physical process may have led to the
apparent cutoff in the observed period distribution.  This
investigation is an extension of our earlier calculations of the
interior structures of weakly eccentric Jovian planets at constant
orbital distances under the influence of interior heating and stellar
irradiation (\cite{peter}; hereafter BLM). In
these previous calculations, the interior heating is imposed to be
constant in time and is uniformly distributed within the planet.  BLM
suggest that this heating flux may be extracted from the tidal
dissipation associated with the synchronization of the planets' spin
with their orbital mean motion and the circularization of eccentric
orbits with small semi-major axes $a\approx 0.04$ AU.  Under these
assumptions, BLM show that Jovian planets can be inflated to
equilibrium sizes larger than two Jupiter radii, given an adequate rate
of interior heating.  In the proximity of their host stars, the Roche
radii of the short-period planets do not greatly exceed their physical
sizes.  Thus, gross inflation of any planet could lead
to tidal disruption and Roche lobe overflow. We investigate this
process and discuss its implications.  

Mass loss induced by Roche lobe overflow may also be important for the
formation of short-period planets.  Shortly after the discovery of the
short-period planet around 51 Peg (\cite{mq95}), we suggested that it
may have been formed at a distance comparable to the orbital radius of
Jupiter (several AU's), after which it migrated toward its host star
(Lin et al. 1996) as a consequence of disk-planet interaction
(\cite{gt80}; \cite{lp86}; \cite{taku}).  The survival of the
short-period planets requires the termination of their orbital
evolution.  Among many scenarios, \cite{tri98} suggest that the size
of $10^6$ yr old Jupiter-mass protoplanets is more than twice that of
Jupiter,  so that they would overflow their Roche lobe if their orbits
decayed to within $\sim 0.03$ AU from their host stars.  \cite{tri98}
further postulate that the lost material flows through the L1 point
and angular momentum is transferred from it to the planet's orbit, which would 
balance the angular momentum loss to the disk  and halt the orbital
decay.

In \S2, we briefly recapitulate the time scale and energy dissipation
rate associated with both synchronization and circularization of
planetary orbits.  We describe the response of the planetary envelope
in \S3 and describe the mass-loss flow pattern associated with
Roche-lobe overflow in \S4.  Based on some analytic approximations, we
illustrate the dynamical evolution of tidally unstable planets in
terms of a sequence of stages in \S5.  
We also present in \S6 the results of numerical calculations of
various models of synchronization and circularization processes.
Finally, we summarize the results and discuss their implications in
\S7.

\section{Tidal interaction between short-period planets and their 
host stars}

The equation of relative motion between a short-period planet with a
mass $M_p$ and its host star with a mass $M_\ast$ and separation $r$ is
\begin{equation}
{d^2 {\bf r} \over d t^2 } = - \frac{G M_{t}}{r^3} {\bf r}
+ \sum_{i=1} ^N {\bf f}_i
\label{eq:eom}
\end{equation}
where $M_{t} = M_p + M_\ast$, and the acceleration ${\bf f}_i$
includes the relativistic potential of the host star, the contribution
from other planets, and the tidal and spin distortions of the star and
the planet (\cite{ml02}). 

\subsection{Synchronization of the planet's spin}
For computational simplicity, we focus our
attention in this paper on the evolution of the spin angular frequency
($\Omega_{p, \ast}$ for the planet and star, respectively), $e$, and
$a$ of one single planet in the limit that its spin and orbital axes
are both parallel with the spin axis of the star. We also assume the
rotational and tidal distortion of the star and the planet is
approximately symmetric about the line separating them so that they do
not significantly affect the orbital evolution.  The only remaining
contribution which can lead to angular momentum transfer and induce
orbital evolution is the acceleration due to the tidal damping within
the planet and the star where
\begin{equation}
{\bf f}_{p,\ast}=
-\left(\frac{3n\,k_{p,\ast}}{Q_{p,\ast}}\right)\left(\frac{M_{\ast,p}}
{M_{p,\ast}}\right)\left(\frac{R_{p,\ast}}{a}\right)^5\left(
\frac{a}{r}\right)^8\left[
3\left(\hat{\bf r}\cdot\dot{\bf r}\right)\hat{\bf r}
+\left(\hat{\bf r}\times\dot{\bf r}-r{\bf \Omega}_{p,\ast}\right)
\times\hat{\bf r}\right],
\label{eq:tf1}
\end{equation} 
where $n = (G M_{t} / a^3)^{1/2}$ is the mean motion of the mutual
orbit of the planet and the star, and $R_{p,\ast}$, $k_{p,\ast}$, and
$Q_{p,\ast}$ are respectively, the size, Love number, and the
$Q$-value, of the star (subscript $\ast$), and the planet (subscript
$p$) (\cite{egg98}).  The above expression is based on the derivation 
for equilibrium tides with a constant tidal lag angle.

The planet's specific orbital angular momentum
${\bf h} = {\bf r \times \dot r}$ changes at a rate
\begin{equation}
{d {\bf h} \over dt} = {\bf r \times f}_t
\end{equation}
where ${\bf f}_t = {\bf f}_p + {\bf f}_\ast$.  The corresponding total
rate of angular momentum transfer from the spin of the planet/star to the 
planet's orbit 
\begin{equation}
\dot J_{p, \ast} = I_{p,\ast} {\bf \dot \Omega}_{p,\ast} 
= - \mu {\bf r \times f}_{p,\ast}
\label{eq:odot}
\end{equation}
where $\mu = M_\ast M_p/M_{t}$ is the reduced mass, and  $I_{p,\ast} =
\alpha_{p,\ast} \epsilon_{p, \ast} M_{p, \ast} R_{p,\ast}^2$ is the
moment of inertia of the planet and the star,  respectively.  The
quantity $\epsilon_{p, \ast}$ represents the fraction of the planet
and star which participates in the tidally induced angular momentum
exchange.  Since gaseous planets have extensive convection zones which
can be fully mixed, $\epsilon_p \simeq 1$.  But, solar type stars have
shallow surface convection zones so that $\epsilon_\ast \sim$ a few
times $10^{-2}$.  The coefficients $\alpha_{p,\ast}$ are determined by
the internal structures  of the planet and the star. The equation of
state appropriate for Jupiter-mass planets can be approximated with a
polytrope with an index 1 (Hubbard 1984), in which case, $\alpha_p =0.26$.

If the planet's/star's spin is aligned with the 
orbit, we find from equations (\ref{eq:tf1}) and (\ref{eq:odot}) that
\begin{equation}
\dot \Omega_{p, \ast} = \left({M_{\ast, p}^2 \over \alpha_{p, \ast}
\epsilon_{p, \ast} M_{p,\ast} M_{t} } \right)
\left({ 9 n \over 2 Q_{p, \ast}^\prime} \right)
\left( {R_{p, \ast} \over a} \right)^3 \left[
f_3 (e) n - f_4 (e) \Omega_{p, \ast} \right] + \dot \omega_{p, \ast }
\label{eq:omedot}
\end{equation}
\begin{equation}
\dot J_{p, \ast} = 
\left( {- \alpha_{p, \ast} \epsilon_{p, \ast}
J_o M_{p, \ast} \over (1 - e^2)^{1/2} M_p} \right)
\left( {R_{p, \ast} \over a} \right)^2 \left( {\dot \Omega_{p, \ast}
\over n} \right)_{\dot \omega_{p, \ast} =0}
= \left({9 J_o M_{\ast, p}^2 \over M_{p} M_{t} } \right)
\left({  f_4 (e) \Omega_{p, \ast} -f_3 (e) n
\over 2 Q_{p, \ast}^\prime (1 - e^2)^{1/2}} \right)
\left( {R_{p, \ast} \over a} \right)^5
\label{eq:jpa}
\end{equation}
where
\begin{equation}
f_3(e) = (1 + {15 \over 2} e^2 + {45 \over 8} e^4 + {5 \over 16} e^6)
(1 - e^2)^{-6},
\end{equation}
\begin{equation}
f_4(e) = (1 + 3 e^2 + {3 \over 8} e^4 ) (1 - e^2)^{-9/2},
\end{equation}
$Q_{p, \ast}^\prime = 3 Q_{p, \ast} / 2 k_{p, \ast}$ is the modified
Q-value (cf. \cite{md99}), $J_o = \mu \vert {\bf h} \vert =
M_p (G M_{t} a (1 - e^2))^{1/2}$ is the total angular momentum of the
planet's orbit, $\dot \omega_{p} (= - \dot I_p \Omega_p /I_p) $ is the
changing rate of planetary spin due to its structural adjustment, and
$\dot \omega_{\ast} $ is the changing rate of spin due to angular
momentum transfer, by processes such as the stellar winds, between the
stars and the external medium (Dobbs-Dixon, Lin, \& Mardling 2002,
hereafter DLM). Although the contribution from $\dot \omega_p$ is
generally negligible, that from $\dot \omega_\ast$ due to the stellar
wind can be important for the evolution of $\Omega_\ast$ and $a$.  In
the limit of small $e$, the rate of change of the planet's angular
frequency is
\begin{equation}
{\dot \Omega}_{p} = {9 n \over 2 \alpha_p Q_p^\prime} 
\left({M_\ast \over M_p} 
\right) \left( {R_p \over a} \right)^3 (n - \Omega_p)
\label{eq:odot1}
\end{equation}
which induces the planet to evolve to a state of synchronization with
$\Omega_p = n$.

\subsection{Circularization of the planet's orbit}
Tidal friction also induces the Runge-Lenz vector, ${\bf e} = {\bf
\dot r \times h} / GM_{t} - {\hat{\bf r}}$, to change (\cite{ml02}) 
at a rate
\begin{equation} 
\frac{d{\bf e}}{dt}=\left( {2({\bf f}_t\cdot\brd)\br
-(\br\cdot\brd){\bf f}_t -({\bf f}_t\cdot\br)\brd \over GM_{t} } \right).
\label{eq:edot}
\end{equation} 
The magnitude and direction of ${\bf e}$ correspond to the planet's
orbital eccentricity and longitude of periapse so that when averaged
over an orbit, the rate of eccentricity change reduces to
\begin{equation}
{d e \over d t} =g_p + g_\ast
\label{eq:edot0_e}
\end{equation}
where
\begin{equation}
g_{p, \ast} = -
{81 e \over 2 Q_{p, \ast}^\prime} \left({M_{\ast, p} \over M_{p, \ast}}
\right) \left( {R_{p, \ast} \over a} \right)^5
\left( f_1(e)n - {11 f_2(e) \Omega_{p, \ast} \over 18} \right),
\label{eq:edot1_e}
\end{equation}
$f_1(e)=(1+15e^2/4+15e^4/8+5e^6/64)/(1-e^2)^{13/2}$, and
$f_2(e)=(1+3e^2/2 +e^4/8)/(1-e^2)^5$.
In the limit of small $e$ and nearly synchronous planetary spin, 
the above expression can be further reduced (\cite{gs66}) to
\begin{equation}
{d e \over d t} = - {63 n e \over 4 Q_p^\prime} \left({M_\ast \over M_p} 
\right) \left( {R_p \over a} \right)^5. 
\label{eq:edot1}
\end{equation}
In the above equation, we only considered the planet's contribution in
${\bf f}$.  When the dissipation of the planets' tide in the star is 
also taken into account, the damping rate is modified from the expression
in equation (\ref{eq:edot1}) by a factor
\begin{equation}
\beta =1 + \left( {2 Q_p^\prime \over 7 Q_\ast^\prime } \right)
\left( {M_p \over M_\ast} \right)^2 \left( {R_\ast \over R_p} \right)^5
\left( 9 - {11 \Omega_\ast \over 2 n} \right).
\end{equation}
For planets with relatively large $R_p$, $\beta \simeq 1$ which
indicates that the star does not significantly modify the $e$
damping rate. If the host star has the same values of $Q^\prime$ and
average internal density as those of the planet, its contribution
would be weaker than that of the planet by a factor of $R_p/
R_\ast$. However, the contribution from a very rapidly spinning host
star, with 
\begin{equation}
\Omega_\ast > \Omega_{pc} \equiv {18 f_1(e) n \over 11 f_2(e)},
\end{equation}
can reduce the total rate of $e$ damping or even induce a net rate of
$e$ excitation with a positive $de/dt$ (DLM).

Equations (\ref{eq:tf1}), (\ref{eq:odot}), and (\ref{eq:edot})
indicate that the evolutionary time scales for $\Omega$ and $e$ are
proportional to the magnitude of $Q_{p,\ast}^\prime$.  These
expressions are derived using equilibrium tidal models with constant
lag angles.  But, in reality, the stellar tidal perturbation is likely
to excite gravity and inertial waves.  These waves which propagate
within the planet are dissipated either through convective turbulence
in its interior or radiative damping near the surface.  The
dissipation of the waves also leads to the deposition of angular
momentum which in general leads to differential rotation.  The nature
of these complex processes are poorly understood and the theoretical
estimates of the $Q_{p,\ast}^\prime$ of Jupiter-mass planets and
solar-type host stars remain uncertain (Zahn 1977, 1989; Goldreich \&
Nicholson 1977, 1989; Ioannou \& Lindzen 1993a,b, 1994; Lubow et
al. 1997; Terquem et al. 1998; Goodman \& Oh 1997).  Some aspects 
of these uncertainties are associated with how and which waves are
excited and dissipated.  Using a prescription for dissipation due 
to convective eddies, Zahn (1989) found an expression  for the tidal 
torque which when scaled with the expression in equation (\ref{eq:tf1}) gives 
a frequency dependence for 
\begin{equation}
Q^\prime _p = {3 \over 4} {M_p \over M_*} {n \, t_{\!f} \over \lambda_2}
\left({a \over R_p}\right)^3 
\label{eq:zahn}
\end{equation}
where $t_f = [M R^2 / L]^{1/3}$ is the tidal dissipation time for a
convective body, and $\lambda_2 \approx 2 \, 10^{-2}$ when it is fully
convective.  But, using a slightly different prescription for the
dissipation in convective eddies, Terquem {\it et al.} (1998) find an
expression for the tidal torque which corresponds to a $Q^\prime _p$
which is independent of $n$.  Even less certain is how the interior of
the gaseous planets and stars may readjust after the angular momentum
deposition may have introduced a differential rotation in their
interior.

The theoretically  determined $Q_\ast$-values using both approaches imply
a slow rate for close binary stars to circularize their orbits, which
is inconsistent with the observations of solar-type binary stars in
stellar clusters of various ages (Mathieu 1994).  Mathieu {\it et al}
(1992) also pointed out that a power-law function ($e/ {\dot e}
P^{13/3}$) provides a close fit to the slope of the observed cutoff
period.  Such an empirical expression would correspond to a $Q'$ value
which is independent of $n$.  Nevertheless, the observations are
equally well fitted with an index of 10/3, and an index of 16/3 cannot
be ruled out.

To carry out a comprehensive analysis on the physical process
associated with the dissipation of the tidal disturbance is beyond the
scope of the present paper.  In view of these uncertainties, we took
the simplest prescription based on an equilibrium tidal model with a
constant lag angle for all components of the tide.  This may not
necessarily be the most reliable model.  All the uncertainties
associated with the physical processes are contained in the $Q^\prime$
values and we shall parametrize our results in terms of it.  For a
fiducial value, we note that the $Q_p ^\prime$-value inferred for
Jupiter from Io's orbital evolution is $5 \times 10^4 < Q_p ^\prime <
2 \times 10^6$ (\cite{yp81}).  With this $Q_p^\prime$-value, orbits of
planets with $M_p$ and $R_p$ comparable to those of Jupiter, and with
a period less than a week, are circularized within the main-sequence
life span of solar-type stars.  If we use equation (\ref{eq:zahn}) to
extrapolate from the observationally determined  values for Jupiter,
$Q^\prime_p \sim 10^8$ which would imply that only systems with
periods less than 3 days would be circularized within a few Gyr.  But,
if all planets are formed {\it in situ} or brought to their present
location shortly after they are formed, their extended radii would
enhance the rate of tidal circularization.  Within the observational
uncertainties, all known extrasolar planets with periods less than a
week and some with periods less than two weeks have circular orbits.

With the theoretical estimate of stellar $Q_\ast^\prime$-values ($\sim
10^8$), the time scale for orbital evolution of Jupiter-mass planets
is longer than the stellar age of a few Gyr unless their orbital
period is less than $\sim 8-9$ hours (\cite{ra96}, Marcy et al.
2000).  But, using the observed circularization time scales as a
calibration, the inferred value of $Q_\ast^\prime$ is $\sim 1.5 \times
10^5$ for the very young binary stars (Lin et al. 1996) and $10^6$ for
the mature binary stars (Terquem et al. 1998).  In this paper, we
shall adopt the observationally inferred values for $Q^\prime _{p,
\ast}$ and discuss the implications of our results if they are 
somewhat larger.  

\subsection{Tidal evolutionary time scales}
With these values of $Q_{p,\ast }^\prime$, the synchronization,
circularization, and migration of the planet's orbit occurs on time
scales ($\tau_\Omega$, $\tau_e$, $\tau_a$) which are much longer than
the orbital period,  and they can be obtained from the time-averaged 
equation of motion.  In the limit of $\beta \simeq 1$, small $e$ and 
nearly synchronous planetary spin,
\begin{eqnarray}
&\tau_{\Omega p}&= {\Omega_p \over | \dot \Omega_p |} = 
0.187 \alpha_p \epsilon_p \left| {1\,{\rm day}/(2\pi /\Omega_p) \over
0.34 (M_\ast/\msun)^{1/2} ( 0.04\,{\rm AU}
/a)^{3/2} - 1\,{\rm day}/(2\pi /\Omega_p)} \right| \nonumber \\
& \quad & \quad \times \left( {Q_p^\prime \over 10^6} \right)
\left( {M_p \over M_J} \right)
\left( {\msun \over M_\ast} \right)^{3/2}
\left( {a \over 0.04\,{\rm AU}} \right)^{9/2}
\left( {R_J \over R_p} \right)^3
\ {\rm Myr},
\label{eq:tauo}
\end{eqnarray}
\begin{equation}
\tau_e ={e\over | de/dt |}
={0.33} \left( {Q_p^\prime \over 10^6} \right)
\left( {M_p\over M_J} \right) \left( {\msun \over M_\ast} \right)^{3/2}
\left( {a \over 0.04\,{\rm AU}} \right)^{13/2}
\left( {R_J \over R_p} \right)^5
\, {\rm Gyrs}. \label{tau_e}
\end{equation}
The above derivation for $\tau_e$ is based on the approximations that  the
planet's spin is nearly synchronized with its mean motion and the tide
raised on the planet by the star provides the dominant contribution to
the $e$ evolution. The magnitude of $\tau_e$ would be
shortened/lengthened by the planet's tide raised on a slowly/rapidly
(relative to $\Omega_{pc}$) spinning star (DLM).  From these equations
we find
\begin{equation}
\tau_{\Omega p} = {7 \alpha_p \epsilon_p \over 2} \left({R_p \over a} 
\right)^2 
\left|{\Omega_p \over \Omega_p - n} \right| \tau_e.
\label{eq:tauoe}
\end{equation}
In the limits that $e < < 1$ and $M_p < < M_\ast$, $\tau_{\Omega p}< < 
\tau_e$, and the nearly-synchronous approximation in equation (\ref{tau_e}) is 
justified self consistently.

Under the action of tidal torque and the conservation of total angular
momentum, $\dot J_o = \dot J_p + \dot J_\ast$.  In the case without
mass transfer between the star and the planet, this exchange also
leads to an evolution of the  semi-major axis at a rate
\begin{equation}
\dot a = a \left( {2 e \dot e \over (1 - e^2)} + {2(\dot J_\ast + \dot
J_p) \over J_o} - { 1 \over \tau_d} \right) 
\label{eq:adot}
\end{equation}
where $\tau_d$ is the usual planetary migration time scale through 
disk-planet interaction. 
Since the planet quickly establishes a state of synchronous rotation,
the magnitude of $\dot J_p$ becomes much smaller than that of $\dot
J_\ast$ and $\tau_e = \vert e/ \dot e \vert < < \tau_{\Omega \ast} <
\vert J_o / \dot J_\ast \vert < \vert J_o / \dot J_p \vert$ so that
$a$ declines together with $e$.  But, after the orbit is circularized,
$a$ and $\dot J_\ast + \dot J_p$ evolve together with the same sign
and the evolution of $a$ is mainly driven by the dominant effect of
the tide raised on the star by the planet (\cite{gs66}, \cite{ml02})
such that
\begin{equation}
\tau_{a} \equiv {a\over \dot a} 
\simeq {Q_\ast ^\prime M_\ast \over 9 M_p} 
{(a/R_\ast)^5 \over (\Omega_\ast -n)}
\simeq \left( {7 n\over \Omega_\ast -n} \right)
\left( {Q_\ast^\prime \over 10^6} \right)
\left( {M_\ast \over \msun} \right)^{1/2}
\left( {M_J \over M_p} \right) 
\left( {a \over 0.04\, {\rm AU}} \right)^{13/2}
\left( {\rsun \over R_\ast} \right)^5
\ {\rm Gyr}
\label{eq:taua}
\end{equation}
in the limit $\tau_d = \infty$.  Provided $Q_\ast ^\prime \sim Q_p
^\prime$ and $\beta \sim 1$, $\tau_a > \tau_e > \tau_\Omega$.  For
$R_\ast=R_\odot$, $\tau_a$ is much larger than 
$\tau_d$ (except for $a <
0.01$ AU), and is also much larger than the relevant time scales
$\tau_R$ and $\tau_e$ for planetary inflation and eccentricity damping
(which are $\sim 1$ Myr, see the discussions in \S3.3, \S5, and the
simulation results in \S6.2) that concern us in this paper.  Thus, we
can neglect the evolution of $a$ due to the tide raised on the star by
the planet today.  But, during the classical T Tauri phase when
$R_\ast \sim 3 R_\odot$, $\Omega_\ast$ is a significant fraction of
the breakup angular frequency, and $Q_\ast \sim 1.5 \times 10^5$, the
magnitude of $\tau_a$ is comparable to that of $ \tau_d
(\sim 10^{5-6}$ yr) at
\begin{equation}
a = a_{\rm halt} \simeq \left( {9 M_p \tau_d \Omega_\ast \over Q^\prime _\ast 
M_\ast} \right)^{1/5} R_\ast \sim 0.03-0.04~{\rm AU}
\label{eq:ahalt}
\end{equation} 
such that protoplanets' migration induced by their tidal
interaction with their protoplanetary disks may be stalled by that with their
host stars (Lin et al. 1996).  Although the magnitude of $\tau_d$ may
differ from its assumed value of $\sim 10^5$ yr in the above estimate,
$a_{\rm halt}$ has a very weak dependence on $\tau_d$.  The value of 
$a_{\rm halt}$ is also a weak function of $Q^\prime _\ast$.  Two orders of
magnitude increase in $Q^\prime _\ast$ would reduce $a_{\rm halt}$ by a
factor of $<3$.

\section{Planet's Structural Adjustment due to its Internal Tidal Energy
Dissipation}

In this section, we consider 1) the effect of tidal dissipation in
increasing the internal temperature of the planet, 2) the necessary
conditions for planetary inflation, 3) the sufficient condition for
Roche lobe overflow.

\subsection{Energy budget}
During the synchronization and circularization processes, kinetic
energy is continually dissipated into heat. From equations
(\ref{eq:adot}) and (\ref{eq:jpa}) we find that both $\dot e$ and
$\dot \Omega_p$ lead to the evolution in $a$ and a corresponding
change in the orbital energy $E_{orb} = - G M_\ast M_p / 2 a$ at a
rate
\begin{equation}
\dot E_{\rm orb} =  - E_{\rm orb} {{\dot a} \over a}.
\end{equation}
In addition, the planet's spin energy $E_{\rm spin} = I_p \Omega_p ^2/2$
evolves at a rate
\begin{equation}
\dot E_{\rm spin} = \dot E_{\rm spin}^R + \dot E_{\rm spin}^\Omega \equiv
\alpha_p M_p R_p \dot R_p \Omega_p ^2 + I_p \Omega_p \dot \Omega_p
\label{eq:espin}
\end{equation}
where the first term on the right hand side represents the planet's
size adjustment (see below), and the second term corresponds to the
changes in the kinetic energy.  From equations (\ref{eq:jpa}) and
(\ref{eq:adot}), we find
\begin{equation}
\dot E_{\rm tide} \equiv \dot E_{\rm orb}  + \dot E_{\rm spin}^\Omega 
= \mu n^2 a^2 \left( { e \dot e \over 1 - e^2} \right)
+  I_p \Omega_p \dot \Omega_p  = \mu ({\bf \dot r} - {\bf 
\Omega_p \times r} ) {\bf \cdot f}_t 
\label{eq:eorig}
\end{equation}
is the total tidal energy dissipation rate within the planet in its
rest frame (\cite{egg98}, \cite{ml02}).  (A similar expression can be
derived for the rate of energy dissipation within the star, but its
effect on the stellar structure is negligible.)  From equations 
(\ref{eq:omedot}), (\ref{eq:edot0_e}), (\ref{eq:edot1_e}), and 
(\ref{eq:eorig}), we find
\begin{equation}
\dot E_{\rm tide} =- \left({9 \mu n a^2 \over 2 Q_p^\prime}
\right) \left({M_\ast \over M_p} \right) \left({R_p \over a} \right)^5
\left[ \Omega_p^2 h_3(e)- 2 n\Omega_p h_4(e) +
n^2 h_5(e) \right] 
\label{eq:etide24},
\end{equation}
where $h_3(e)=(1+3e^2 + 3e^4/8) (1-e^2)^{-9/2}$,
$h_4(e)=(1 + 15 e^2/2 +45 e^4/8 + 5 e^6/16) (1 - e^2)^{-6}$, and 
$h_5(e)=(1+31e^2/2+255e^4/8+185e^6/16+25e^8/64) (1 - e^2)^{-15/2}$.

For small $e$'s, the leading terms of $e^2$ and $\Omega_p/n
-1$ in equation (\ref{eq:etide24}) are reduced to
\begin{equation} 
\dot E_{\rm tide} = -{9 \mu a^2 n^3 \over 2 Q_p ^\prime}
\left({M_\ast \over M_p} \right) \left({R_p \over a} \right)^5
\left[ \left({\Omega_p \over n} -1 \right)^2 + e^2 \left( {15 \Omega_p ^2 
\over 2 n^2} - {27 \Omega_p \over n} + 23 \right) \right]
\label{eq:etide}
\end{equation}
The first term in the square bracket on the right hand side of
the above equation represents the energy dissipation due to the
synchronization of the planet's spin.  If planets are formed as rapid
rotators, similar to Jupiter and Saturn, equation (\ref{eq:etide}) reduces to
\begin{equation}
\dot E_{\rm tide} \simeq \dot E_{\rm t\Omega} 
\equiv - {I_p (\Omega_p -n) \Omega_p \over  \tau_\Omega}.  
\label{eq:etide2}
\end{equation}
When synchronization is established the terms inside the last bracket
are reduced to $7 e^2/2$ and equation  (\ref{eq:etide}) may be approximated as
(c.f. \cite{gs66})
\begin{equation}
\dot E_{\rm tide} \simeq \dot E_{\rm te} 
\equiv - {e^2 G M_\ast M_p \over a \tau_e}.  
\label{eq:etide3}
\end{equation}

We now consider the effect of tidal dissipation on the total energy
budget of the system.  In response to this source of energy, the
planet adjusts its internal structure to evolve toward an energy
equilibrium in which the rate of change of the planet's total energy
is given by
\begin{equation}
\dot E_{\rm tot} =  \dot E_{\rm bind} + \dot E_{\rm spin} = - {\cal L}
\label{eq:etot}
\end{equation}
where ${\cal L}$ is the rate of loss of energy from the system via
radiation from the planet's surface.  In this derivation, we assume
that the planet maintains a state of hydrostatic equilibrium.
The tidal energy dissipation rate is
contained in $\dot E_{\rm tot}$ in the following manner.  The binding
energy has three contributing components 
\begin{equation}
{E}_{\rm bind}=-q_\ast \frac{G M_\ast^2}{R_\ast(t)} 
-q_p\frac{G M_p^2}{R_p(t)} + E_{\rm orb} 
\label{eq:ebind}
\end{equation}
where $q_{\ast, p}=3/(5-n_{\ast, p})$ and $n_{\ast, p}$ is the
polytropic index for the star and planet respectively. For a polytrope
of index 1, $q_{\ast, p} =3/4$ and for a uniform distribution of
density, $q_{\ast, p}=3/5$.

Neglecting any changes in the gravitational energy of the host star, 
\begin{equation}
\dot {E}_{\rm bind}=\frac{q_p G M_p^2}{ R_p(t)}\frac{\dot R_p}{R_p} 
+ \frac{G M_p M_\ast}{2 a(t)}\frac{\dot a}{a}.  
\end{equation}
Note that $M_p$ in the above equation can also be a function of time
when the size of planet is inflated up to its Roche radius and therefore
mass loss occurs.
Similarly, the spin energy can be separated into two pieces (see eq. 
\ref{eq:espin}).  In principle, rotation may also affect the binding
energy of the planet.  But from equations (\ref{eq:ebind}) and
(\ref{eq:espin}), we find the energy ratio
\begin{equation}
{E_{\rm spin} \over |E_{\rm bind}|} = {\alpha_p \over 6q_p} \left({R_p 
\over R_L} \right)^3 \left( {\Omega_p \over n} \right)^2 (1-e)^3
\label{eq:espbd}
\end{equation} where
\begin{equation}
R_L = \left({M_p \over 3M_\ast} \right)^{1/3} a(1-e) \approx 5.74 \left( 
{M_p \over M_J} \right)^{1/3} \left( {\msun \over M_\ast} \right)^{1/3}
\left( {a \over 0.04 {\rm AU}} \right) (1-e) R_J, \label{r_Roche}
\label{eq:roche}
\end{equation}
is the planet's Roche radius and  $M_J$ and $R_J$ are Jupiter's mass and
radius, respectively.  After the planet attains synchronization within
its Roche radius, its rotational energy is relatively small compared
with its binding energy.  Since $E_{\rm spin} < < E_{\rm bind}$, the
effects of Coriolis and centrifugal force are negligible in the
calculation of the planet's internal structure.  In the above
expression for $R_L$, we introduce a factor of $(1-e)$ to represent
the tidal radius at periastron where the stellar tidal effect is at an
maximum.

\subsection{Tidal inflation instability}
Tidal dissipation within the planet leads to changes in $a$ and
$\Omega_p$, and we have already defined the  rate $\dot E_{\rm tide}$ in
equation (\ref{eq:eorig}).  This process also leads to the internal
heating of the planet at a rate $- \dot E_{\rm tide}$. From the above
equations, we find that the planet's size would change at a rate
\begin{equation}
\dot{R}_p=\frac{-{\cal L}-\dot{E}_{\rm tide}}
{q_p G M_p^2/ R_p^2+\alpha_p M_p R_p \Omega_p^2} 
\label{rd}
\end{equation}
provided the adjustment would not directly modify $\alpha_p$ (\cite{ml02}).

In a thermal equilibrium, $\dot R_p =0$ and equation  (\ref{rd}) reduces to 
\begin{equation}
{\cal L} = {\cal L}_e \equiv -\dot E_{\rm tide} 
\label{eq:thermal}
\end{equation} 
such that the surface radiative flux is balanced by internal energy
dissipation. In response to a small perturbation $\delta R_p$ 
from its  equilibrium radius 
\begin{equation}
R_e \equiv R_p ({\cal{ L}} = -\dot E_{\rm tide})
\label{eq:Re}
\end{equation}
and in the case of  $\tau_e >> \tau_R$ ({\it i.e.} $e$ can be
roughly treated as a constant with time for this small perturbation),
a planet adjusts its  radius at a rate
\begin{equation}
\dot R_p  = \left({9 \mu n a^2 \over 2 Q_p^\prime}
\right) \left({M_\ast \over M_p} \right) \left({R_e \over a} \right)^5
\left( { \Omega_p^2 h_3(e)- 2 n\Omega_p h_4(e) +
n^2 h_5(e) \over q_p G M_p^2/ R_p^2+\alpha_p M_p R_p \Omega_p^2 }\right) 
\left( 5 - \gamma \right) \left({\delta R_p \over R_e} \right)  
\end{equation}
where 
\begin{equation}
\gamma \equiv \left( {\partial {\rm ln} {\cal L} \over \partial 
{\rm ln}  R_p} \right)_{R_p = R_e} .
\label{eq:gamma}
\end{equation}
Note that $\dot R_p$ and $\delta R_p$ would have
opposite signs and the thermal equilibrium would be stable if
$\gamma >5$.  But if $\gamma <5$, small increases in $R_p$ would lead
to runaway inflation.

In general $\cal L$ is a function of $M_p$, $R_p$, and the existence
of the core.  Based on three sets of equilibrium
models, BLM obtained an approximate $R_e-{\cal L}$ relations in which
\begin{equation}
{\rm log} {R_e \over R_\odot} = A(M_p) + B(M_p) {\rm log} 
{{\cal L} \over L_\odot} + C(M_p) \left( {\rm log} 
{{\cal L} \over L_\odot} \right)^2
\label{eq:equalL}
\end{equation}
where, in the range $10^{-8} L_\odot < {\cal L} < 10^{-5} L_\odot$,
$A=(3.11, 1.012, -0.269)$, $B=(1.011, 0.4689, 0.1631)$, and
$C=(0.0642, 0.0289, 0.00978)$ for $M_p = (0.63, 1, 8) M_J$
respectively.  In the high luminosity limit, $R_e$ becomes a more
rapidly increasing function of ${\cal L}$ (with a decreasing $\gamma$
magnitude) as degeneracy and non-ideal effects in the equation of
state are partially lifted for $R_p > 2 R_J$.  Based on the results of
a limited number of models computed by BLM for $M_p = 0.63$ and $1
M_J$ planets around HD 209458 and Ups And
\footnote{The semi-major axis of HD209458b is actually about 0.045
AU. The only difference modeled by BLM for giant planets located at
different $a$ is the degree of stellar irradiation. However, BLM show
that inflation of giant planets due to stellar irradiation is
negligible compared with tidal inflation. Therefore the estimate
performed in the text section based upon replacing 0.045 AU by 0.04 AU
should serve as a  good approximation.}, we find that $\gamma$ crosses the
critical value of 5 when $R_p = R_c \sim 2 R_J$ in both cases and
${\cal L}_e = {\cal L}_c = (10^{-6}, 10^{-5.5} L_\odot)$ for $M_p =
(0.63, 1 M_J$) respectively.

Although $\gamma$ continues to decline for $R_p > R_c$, we introduce
a power-law approximation   
\begin{equation} 
{{\cal L}_e \over {\cal L}_o }
\simeq \left( {R_p \over 2 R_J} \right)^\gamma.
\label{eq:cale}
\end{equation}  
for a limited range of $R_p$. 
In the range of $R_p = 2-5 R_J$, we fit the
results of BLM with a normalization coefficient ${\cal L}_o =10^{-6}
L_\odot$ and $\gamma= 3$ for a $M_p = 0.63 M_J$ planet. This
prescription is chosen for computational convenience and consistency,
{\it i.e.}  ${\cal L}_e = {\cal L}_c$ for $R_p = R_c \sim 2 R_J$.  In
\S5, we present additional numerical models which also indicate
$\gamma < 5$ for $R_p > 2 R_J$ for other values of $M_p$.  

In this limit, the thermal equilibrium is linearly unstable to
tidal dissipation. A small
enlargement in the planet's size from $R_e$ would cause tidal
dissipation to increase more rapidly than the radiative losses on its
surface. However, $e$ decays in the meantime.
The question of whether the planet's $R_p$ would finally
reach the Roche radius depends on the magnitude of $e$:
there exists a threshold $e$ beyond which  the planet is
able to be inflated to its Roche radius, meaning that
this process relies on a nonlinear instability.
Similarly, a small reduction in $R_p$
from $R_e$ would lead to rapid        cooling,  and the planet would
contract until $R_p$ is reduced to $R_c$.
The characteristic contraction time scale at $R_p = R_c$ is
\begin{equation}
\tau_c = { q_p G M_p^2 \over R_c {\cal L}_c} = 1.6 \times 10^8 q_p
\left( {10^{-6} L_\odot \over {\cal L}_c} \right) \left( {2 R_J \over
R_p } \right) \left( { M_p \over M_J} \right)^2 {\rm yr}
\label{eq:taucool}.
\end{equation}

\subsection{Necessary and sufficient condition for runaway 
planetary inflation} The above criterion for tidal inflation
instability ($\gamma < 5$) is applied to a state of thermal
equilibrium.  In general, the planets may not be in such an
equilibrium, and during the evolution toward it, tidal dissipation may
modify their orbital properties. Therefore, it is essential to
identify 1) the sufficient kinematic properties which would enable
planets to inflate before their eccentricity is damped out and 2) the
necessary initial conditions which may lead to an equilibrium radius
$R_e$ which exceeds a critical radius $R_c (\sim 2 R_J$) for the onset
of tidal inflation instability.  In the determination of the stability
criteria, we must take into account the dynamical evolution of the
planets' orbits.  

Prior to reaching an equilibrium value of $R_e$, the planet's radius
$R_p$ adjusts at a rate in accordance with equation (\ref{rd}).  Since
$E_{\rm spin} < E_{\rm bind}$, this equation implies that 
$R_p$ inflates on a time scale
\begin{equation}
\tau_{\rm R} = {R_p \over \dot R_p} \simeq -{q_p G M_p^2 \over 
\dot E_{\rm tide} R_p}
\label{eq:taur}
\end{equation}
in the limit that the tidal heating greatly exceeds the radiative 
cooling.  

Note that equation (\ref{rd}) is based on the implicit assumption that
$\dot R_p$ is uniquely determined by a given set of $R_p$, $M_p$, and
$\dot E_{\rm tide}$ regardless of the thermal history of the planet's
interior.  In principle, the evolution of stratification of planets
due to thermal imbalance in a state of quasi-hydrostatic equilibrium
should be a function of the heating mechanism.  But, gas inside
Jupiter-mass giant planets is partially degenerate and non-ideal such that the
total pressure is insensitive to temperature changes.  BLM presented
several models with different dissipation prescriptions.  The
results of all their models show that a large fraction of the energy
dissipated within the planet leads to temperature increases without
significantly modifying the pressure and thermal stratification of the
convective interior of the planet. Only the tenuous radiative zone in
the outermost 10\% of the planet's radius responds to heating like an
ideal gas.  Thus, tidal dissipation only affects the boundary
condition at the interface between the convective and radiative zones
as well as the structure of the radiative zone,  and our approximation
is justified.

During the process of synchronization, although the total amount of
energy available ($E_{\rm spin}$) is limited, it is dissipated on a
rapid time scale of $\tau_\Omega$ such that $\dot E_{\rm tide} \simeq
\dot E_{\rm t\Omega}$.  From equations (\ref{eq:etide2}) and
(\ref{eq:taur}), we find
\begin{equation}
\tau_{\rm R} \simeq \tau_{{\rm R} \Omega} \simeq {\Omega_{kp}^2 
\tau_\Omega \over \Omega_p |\Omega_p -n|}  \left({ q_p \over  \alpha_p } 
\right) 
\label{eq:tauro}
\end{equation}
where $\Omega_{kp}$ is the breakup spin frequency of the planet, and
$\tau_{{\rm R} \Omega}$ is the planetary-inflation time scale in the
limit where only the effect of synchronization is considered.  Since
$\Omega_p < \Omega_{kp}$ and $\alpha_p$ is relatively small,
$\tau_{{\rm R} \Omega} > \tau_\Omega$ and the planet does not have
time to inflate in size before its spin become synchronized with its
mean motion.

Energy is also dissipated within the planet during $e$ damping.  In
order for a planet to be inflated by $e$ damping against its own
gravitational binding energy, $e$ should decay on a longer time scale
($\tau_{\rm e}$) than the thermal time scale associated with inflation
($\tau_{\rm R}$); that is, the system should not run out of its
dissipation energy source before it overflows its Roche radius. From
equations(\ref{eq:etide3}) and (\ref{eq:taur}), we find that, in the limit
that tidal dissipation mainly leads to circularization,
\begin{equation}
\tau_{\rm R} \simeq \tau_{{\rm R} e} \simeq q_p\left({M_p \over M_\ast} \right)
\left({ a \over R_p } \right) \left({ \tau_e \over  \beta e^2} \right) 
\simeq 3q_p \left({ R_L \tau_e \over  R_p \beta} \right) \left({R_L \over a e} 
\right)^2 = {e_R^2 \tau_e \over e^2}.
\label{eq:taur1}
\end{equation}
The necessary condition for planetary inflation ($\tau_{\rm R}
<\tau_e$) is satisfied in the limit that its $e$ is greater than a
critical value
\begin{equation}
e > e_R \simeq \left( {q_p \over \beta} 
{M_p \over M_\ast} {a \over R_p} \right)^{1/2}
= 0.18 \left( {q_p  \over 0.75 \beta}
{ M_p \over M_J} {M_\odot \over M_\ast} {2 R_J \over R_p} 
{a \over 0.04 {\rm AU}} \right)^{1/2}.
\label{eq:er}
\end{equation}
In the above expression, $e_R$  decreases with $R_p$ so that the most
stringent requirement for a protoplanet to inflate is set prior to its
expansion.  If short-period planets migrated to their present location
within a few Myr after their dynamical accretion of ambient gas in the
protostellar disk has ceased, their initial $R_p$ would be $\sim 2
R_J$ (\cite{pol96}).  In equation (\ref{eq:er}), we use a fiducial value of
$2 R_J$ for $R_p$ which is also comparable to the value of
$R_c$. During the same early epoch, main sequence G dwarf stars are
rapid rotators and have relatively small $Q_\ast^\prime$ so that for
relatively large $R_p$, $\beta$ is slightly less than unity.  Note that
in equation (\ref{eq:er}) $e_R$ is independent of $Q_p^\prime$ such that it
is not subject to the uncertainties in the tidal dissipation rate. 

Short-period planets with initial $ e >e_R$ would inflate until they
reached  a thermal equilibrium with a radius $R_p = R_e$.  Those planets
with $e > > e_R$ would retain their initial $e$ as they approached  the
thermal equilibrium.  But those planets with an initial $e$ marginally
larger than $e_R$ would undergo a substantial $e$ damping prior to
reaching the thermal equilibrium. Thus, it is possible for some
planets to initially satisfy the necessary condition for tidal
inflation, and become stabilized after their $e$ is substantially
reduced.  Thus, the instability criterion in equation (\ref{eq:er}) should be
regarded in a qualitative sense. 

Equation (\ref{eq:Re}) indicates that $R_e$ is a function of $\dot
E_{\rm tide}$.  Assuming the planet has already attained spin-orbit
synchronization (so that $\dot E_{\rm tide} \sim \dot E_{\rm te}$), we
find from equations(\ref{tau_e}), (\ref{eq:cale}), and (\ref{eq:Re}) that
\begin{equation}
{R_e \over R_s } = \left( {e_L (R_s) \over e} \right)^{2 \over 5 - \gamma}
\label{eq:Re1}
\end{equation}
where $R_s$ is a fiducial length scale and
\begin{eqnarray}
&e_L (R_s) &= \left({ a \tau_e (R_e=R_s) {\cal L}_e (R_e=R_s) \over G M_\ast 
M_p}\right)^{1/2} \nonumber \\
& \quad & \simeq 0.056
\left( {Q_p ^\prime \over        10^6}  \right)^{1/2}
\left( {M_\odot \over M_\ast         }  \right)^{5/4} 
\left( {a       \over 0.04   {\rm AU}}  \right)^{15/4} 
\left( {{\cal L}_o \over 10^{-6} L_\odot}  \right)^{1/2}
\left( {2 R_J   \over             R_s}  \right)^{(5 - \gamma)/2} \nonumber \\
& \quad & \simeq 0.06 
\left( {Q_p ^\prime \over        10^6}  \right)^{1/2}
\left( {P           \over3 {\rm days}}  \right)^{5/2} 
\left( {{\cal L}_o \over 10^{-6} L_\odot}  \right)^{1/2}
\left( {2 R_J   \over             R_s}  \right)^{(5 - \gamma)/2}
\label{eq:eL}
\end{eqnarray}
where we adopt the approximation in equation (\ref{eq:cale}) for ${\cal L}$
and scale $R_s$ with $2 R_J$.  Note that although $M_p$ does not
appear directly in equation (\ref{eq:eL}), both $e_L$ and ${\cal L}_o$ are
slowly increasing functions of $M_p$.

In the previous section, we showed  that a planet would adjust to a
stable equilibrium size if $R_e < R_c (\sim 2 R_J)$.  But, for $R_e >
R_c$, $\gamma  <5$, and the
planet is unstable such that it would undergo runaway inflation if its
current size $R_p < R_e$ or contraction to $R_s$ if $R_p > R_e$.  In
principle, we should set $R_s = R_p$ in equations (\ref{eq:Re1}) and
(\ref{eq:eL}).  But, the critical $e_L$ should not be a decreasing
function of $R_s$ as indicated in equation (\ref{eq:eL}).  For example, the
Roche radius of a $0.63 M_J$ planet, located at $a = 0.04$AU around a
1.1 $M_\odot$ star (HD 209458), is $R_L = 4.92 R_J$.  An extrapolation
from Fig. 3 in BLM shows that an internal heating rate of at least
$\approx 10^{-4.9}\,\Lsun$ is required for a coreless $M_p = 0.63 M_J$
planet to attain $R_e = R_L$ in this case.  Nevertheless, the critical
eccentricity $e_L$ for $R_e$ to reach $R_L$ is less than half that
needed for $R_e$ to reach $\sim 2 R_J$ ({\it i.e.} $<0.5 e_L$) because
the efficiency of tidal dissipation increases rapidly with $R_p$.
Thus, analogous to the planet's $R_p$ dependence in $e_R$, the most
stringent requirement for a protoplanet to inflate is set prior to its
expansion rather than the final post-inflation value of $R_L$.  In the
limit that the planet's initial radius $R_i < R_e$, the appropriate
sufficient condition for the onset of runaway inflation is
\begin{equation}
e>e_L (R_s = R_c).
\label{eq:elstable}
\end{equation}  
Once again, note that $e$ may be damp during the course of the planet's
size adjustment, especially for those systems with $e (R_i)$ marginally
larger than $e_R (R_i)$.  In these cases, the value of the $e$ in the
above expression should be $e (R_c)$ (see further discussions in \S6).

The power-law approximation in equation  (\ref{eq:cale}) and its
extrapolation in equation  (\ref{eq:eL}) are only appropriate
for $R_p$ in the range of $1-3 R_c$ where $\gamma < 5$.  In the limit
$R_s = R_c$, $\gamma=5$, ${\cal L}_e = {\cal L}_c$, and 
\begin{eqnarray}
&e_L (R_s=R_c) & 
\simeq 0.056
\left( {Q_p ^\prime \over        10^6}  \right)^{1/2}
\left( {M_\odot \over M_\ast         }  \right)^{5/4} 
\left( {a       \over 0.04   {\rm AU}}  \right)^{15/4} 
\left( {{\cal L}_c \over 10^{-6} L_\odot}  \right)^{1/2} \nonumber \\ 
& \quad & \simeq 0.06 
\left( {Q_p ^\prime \over        10^6}  \right)^{1/2}
\left( {P           \over3 {\rm days}}  \right)^{5/2} 
\left( {{\cal L}_c \over 10^{-6} L_\odot}  \right)^{1/2}. 
\label{eq:eeel}
\end{eqnarray}

\subsection{The onset of planets' inflation to their
Roche lobes}

If a planet migrated to the vicinity of its  host star with an
initial radius $R_i < R_c$ and with  $e$  larger than both $e_L$
in equation (\ref{eq:eeel}) and $e_R$ in equation  (\ref{eq:er}), it  
would undergo runaway tidal inflation. For $a > a_{RL}$ where
\begin{equation}
a_{RL} = 0.06 \left( {q_p / \beta \over 0.75} {10^6 \over Q_p ^\prime}
{M_p \over M_J} {10^{-6} L_\odot \over {\cal L}_o} \right)^{2/13}
\left({M_\ast \over M_\odot} \right)^{3/13} \left( {2 R_J \over R_p}
\right)^{2 \gamma -8 \over 13} {\rm AU},
\label{eq:aRL}
\end{equation}
$e_R < e_L$ so that the condition for planetary inflation is $e >
e_L$.  But, a planet with $a < a_{RL}$, can only inflate if $e > e_R$.
For a $M_p = 0.63 M_J$ planet ($\gamma=3$ and ${\cal L}_o = 10^{-6}
L_\odot$), $a_{RL}$ increases with $R_p$.  Although it is possible for
$e_L (R_c) > e_R(R_c)$ and $e_L(R_L) < e_R (R_L)$ at some locations,
the runaway tidal inflation following the onset of the thermal
instability always leads the planet to overflow its Roche lobe because
both $e_L (R_c) > e_L (R_L)$ and $e_R (R_c) > e_R(R_L)$ such that the
sufficient and necessary conditions for runaway inflation become more
easily satisfied during the planet's expansion. Note that the
magnitude of $a_{RL}$ is very weakly dependent on that of $Q^\prime
_p$.  Two orders of magnitude uncertainty in $Q^\prime _p$ would
correspond to a factor of two difference in $a_{RL}$. Thus, our 
results are quite robust despite the uncertainties in the magnitude of
$Q^\prime _p$.

Equation (\ref{eq:er}) indicates that $e_R \propto a^{1/2}$ and it reduces to
$\sim 0.06$ near the surface of a star.  However, even in the limit
that $e$ is larger than both $e_R$ and $e_L$, the planet's $R_p$ 
is limited by $R_L$ which is smaller than $R_c$ if 
\begin{equation}
a < a_c =0.0133 \left( { M_J \over M_p } \right)^{1/3} \left( {M_\ast 
\over M_\odot} \right)^{1/3} \left( {R_c \over 2 R_J} \right) {\rm AU}.
\end{equation}
Inside $a_c$, planets cannot expand into the unstable range.
Nevertheless, $R_e>R_c$ (and therefore $R_L$) for $e>e_L (R_c)$ which
is very small at such close range (see eq. \ref{eq:eL}).  In this
limit, planets with $e>e_R$ (so their $e$-value may be maintained )
would overflow their Roche radius for all values of $\beta$.

The metallicity of many planet-bearing stars is higher than the solar
value.  It has been suggested that the outer convective zone of some
of these stars may be contaminated by the consumption of one or more
planets which may have undergone orbital decay due to the star-planet
tidal interaction (Lin 1997).  Previous models for such a process
focus on the disintegration of planets inside the envelope of their
host stars (Sandquist et al. 1998, 2002).  The present results
suggest that with a modest $e$-value, short-period planets may be
disrupted during their tidal orbital decay 
prior to entering the envelope of their host stars.

A planet with $a > a_{RL}$ and $R_i < R_c$ would undergo runaway
inflation if  $e>e_L(R_c)$.  The expression in equation(\ref{eq:eeel})
also indicates that $e_L \propto a^{15/4}$ so that a  planet would
inflate and fill its Roche radius provided its initial $a$ is less
than $a_{eL}$ where
\begin{equation}
a_{eL} =0.086 e^{4/15} \left({ 10^6 \over Q^\prime_p} \right)^{2/15} 
\left( { R_c \over 2R_J} \right)^{2/3} \left( {10^{-6} L_\odot
\over {\cal L}_c} \right)^{2/15} {\rm AU}
\end{equation} 
or initial $P$ is less than $P_{eL}$ where
\begin{equation}
P_{eL} = 9.24 e^{4/15} \left( { 10^6 \over Q^\prime_p} \right)^{1/5} 
\left( { R_c \over 2R_J} \right) 
\left( { 10^{-6} L_\odot \over {\cal L}_c} \right)^{1/5} {\rm days}.  
\label{eq:peL}
\end{equation}
Over the possible range of Jupiter's Q-value $5 \times 10^4 < Q_p
^\prime < 2 \times 10^6$, $R_c (\simeq 2 R_J)$, and ${\cal L}_c
(\simeq 10^{-6} L_\odot$ for the $M_p = 0.63 M_\odot$ planet), tidal
inflation from $R_i < R_c$ to Roche lobe overflow becomes unattainable
for $a>0.08-0.13$ AU or $P>8-17$ days.  For these values of $a$,
$M_p$, $R_c$, and $\beta \sim 1$, equations (\ref{tau_e}) and
(\ref{eq:taur}) imply $\tau_{R} \sim \tau_e <$ a few Myrs.

It is possible, though less likely, that protoplanets may be brought
to the vicinity of their stars while they are still in the accretion
phase so that their initial $R_i$ is larger than $R_c$. In this case,
the planets are already in a thermally unstable region.  These planets
would undergo runaway inflation if their $R_e > R_i$ and contraction
to $R_c$ if their $R_e < R_i$.  A direct comparison between $R_e$ and
$R_i$ can be made with equations  (\ref{eq:Re1}) and (\ref{eq:eL}) by setting
$R_s = R_i$ such that
\begin{equation}
{R_e \over R_i } = \left( {e_L (R_i) \over e} \right)^{2 \over 5 - \gamma}
\label{eq:Rei}
\end{equation}
where 
\begin{equation}
e_L (R_i) = \left( {R_c \over R_i } \right)^{(5-\gamma)/2} e_L (R_c) <
e_L (R_c).
\end{equation}
For $a > a_{RL}$, runaway inflation from $R_i (>R_c)$ would occur if
$e > e_L (R_i)$.  

In the above derivation of $e_L$, we explicitly assumed that the  planet's
spin is nearly synchronized with its orbit.  Equation  (\ref{eq:etide})
clearly shows that a significant departure from the $\Omega=n$
synchronization state would lead to  an additional source of tidal
dissipation of energy.  During its inflation, the planet's spin
frequency would decrease with a rate $ \dot \Omega_{pc} = -2 \Omega_p
\dot R_p / R_p = -2 \Omega_p/ \tau_{R}$ if its spin angular momentum
is conserved.  Nevertheless, quasi-synchronization would be maintained
if $ \dot \Omega_{pc}$ is balanced by the tidally induced ${\dot
\Omega}_p$ in equation (\ref{eq:odot1}), {\it i.e.}
\begin{equation}
\tau_\Omega \sim \tau_{R}/2.
\label{eq:tauoth}
\end{equation}  
Under the assumption that most of the energy is dissipated through
circularization (an assumption to be satisfied self consistently
below), equations (\ref{eq:tauoe}), (\ref{eq:taur1}), and (\ref{eq:tauoth})
imply that the planet's spin attains an equilibrium with
\begin{equation}
{\Omega_p - n \over \Omega_p} \simeq {14 e^2 \over 3q_p} 
\left({R_p \over R_L} \right)^3.
\end{equation}  
With such a small departure of $\Omega_p$ from $n$, energy dissipation
associated with the circularization process continues to provide the
dominant heat source as shown in equation (\ref{eq:etide}).

\subsection{Planets' initial eccentricity}
Extrasolar planets with $a \gtrsim$ 0.1 AU have a nearly uniform $e$
distribution between 0-0.7.  All of the dozen extrasolar planets with
period between 3-7 days and most planets with $a \lesssim$ 0.1 AU have
negligible $e$'s. This dichotomy is probably due to the tidal
circularization of short-period planets' orbits during the main
sequence life span of their host stars. All the observed short-period
planets have $a > 0.04$AU. Although some of these planets may have
undergone tidal inflation if their initial $e>e_r$ (see eq. \ref{eq:er}), 
but outside $\sim 0.08-0.13$ AU, $e_L$ rapidly increases above unity 
(see eq. \ref{eq:eL}) so that the rate of tidal dissipation is so weak 
that the planets' thermal equilibrium size is always less than their 
Roche radius.

Interior to 0.04 AU, however, a planet would be inflated to overflow
its Roche radius provided $\tau_{r} < \tau_e$ or equivalently $e >
e_r$ which is $< 0.18$ (see eq. \ref{eq:er}).  We note that if the
extrasolar planets with $a \gtrsim$ 0.1 AU retain their {\it ab
initio} $e$ distribution, most planets, including those with
$a<0.04$AU, would have initial $e>e_r$.

In addition to the extrapolation of the $e$ distribution from the
long-period planets, there are several theoretical reasons to expect
the initial $e$ of short-period planets to be much larger than their
present values.  These short-period planets are most likely to have
formed at much larger distances from their host stars.  Long-term
dynamical instability may induce the mature planets to undergo close
encounters with each other, causing some planets to be scattered to
the close proximity of their host stars with large $e$'s (\cite{rf96},
Ford 1996, \cite{wm96}, \cite{li97}). In
comparison with the value of $e_r$, these planets would satisfy the
necessary criterion for inflation when their $a$ is reduced (along
with their $e$) by the circularization process (Rasio  et al.
1997) to a magnitude $<$0.04AU.

Short-period planets may also have been brought to the proximity of
their host stars during their formation stage as a consequence of the
planet-disk tidal interaction (Lin et al.  1996) through wave
excitation at the planets' corotation and Lindblad resonances.  Energy
and angular momentum transfers via the planets' corotation resonances
damp their $e$'s whereas those through their Lindblad resonances
excite their $e$'s (\cite{gt80}).  Both processes induce gap formation
in the disk near the planets' orbit.  Massive planets with $M_p > 0.01
M_\ast$ clear a wide gap such that regions containing all the corotation
resonances are void of gas.  The interaction between the disk gas and
these massive planets through the Lindblad resonances excites $e$ to
several times the aspect ratio of the disk (\cite{art92},
\cite{pap01}, \cite{gs02}) which is typically $> e_r$.

In contrast, modest-mass planets with $M_p < 0.01 M_\ast$ open narrow
gaps, and residual disk gas may still occupy their corotation
resonances close to their orbits.  The dominance of the corotation
over the Lindblad resonances induces $e$ damping to negligible values.
However, short-period protoplanets' $e$'s may also be excited by the
processes which halted their orbital migration.  For example, some
young stellar objects are observed to have strong magnetic fields
(\cite{jk99}) and their magnetospheres may clear an
inner cavity in their surrounding protostellar disks (Konigl 1991; Shu
et al. 1994).  As they enter into the cavity and retreat inwards from
the disk's inner edge, the planets' corotation resonances evolve into
the tenuous regions prior to their more distant Lindblad resonances.
Thus, during the decline of the planet-disk tidal interaction, the
$e$-damping process diminishes more rapidly than the $e$-excitation
process.  Eventually, the planets' inward migration is terminated when
all of their lowest-order Lindblad resonances have retreated into the
cavity.  But, during the transition stage, the planets' $e$'s  may grow
to be comparable to the fractional difference between the corotation
and Lindblad resonances which is $ \gapprox e_r$ inside 0.04 AU.

Planets' migration may also be stalled if their loss of angular
momentum to the disk is compensated by a transfer of angular momentum
from the spin of their rapidly rotating host stars through their
mutual tidal interaction.  This process is effective in the limit that
$a$ is sufficiently small ($< a_{\rm halt}$) for the stellar tidal
force to be intense and that the stellar spin frequency
$\Omega_\ast>n$ (see eq. \ref{eq:ahalt}).  Fast $\Omega_\ast$ are
observed among young stellar objects, with some reaching the stellar
breakup values (\cite{sta99}).  In equation (\ref{eq:edot1}),
the star's contribution to $\dot e$ is neglected.  However, the
stellar tidal effect would excite the planets' $e$ if $\beta < 0$ or
equivalently,
\begin{equation}
\Omega_\ast > \left[ {7\over 11} \left( {Q^\prime_\ast \over Q^\prime_p} 
\right) \left({M_\ast \over M_p} \right)^2 \left( {R_p \over R_\ast } 
\right)^5  + {18 \over 11} \right] n 
\end{equation}
which $\simeq 8 n$ in the limit that $Q^\prime_s = Q^\prime_p$ and the
average density of the planet equal to that of the star (DLM).

Another mechanism to stop planets' migration is through the rapid
depletion of the disk which would quench their tidal interaction.  Signs
for additional planets are found around more than half of all the host
stars with known planets.  The $e$'s of multiple planets
around the same host star are modulated as they exchange angular
momentum through secular interaction. During the depletion of the
disk, the procession frequencies of the planets also evolve with the
gravitational potential of the depleting disk.  This evolution leads
to a sweeping secular resonance (\cite{ward81}) which may also induce 
the excitation of eccentricity for some planets (\cite{na02}).

\section{Mass Loss from Inflated Planets}
Planets with $e > e_r$ inflate before their orbits are circularized.
If their $e>e_L$, the expansion of their envelope would continue until
it reached  the Roche radius.  In this section, we consider the flow
pattern associated with the mass loss process.
 
\subsection{Density scale height near the Roche lobe}
In a quasi-hydrostatic equilibrium, the planet's surface is determined
by the equipotential surfaces.  For planets with circular orbits, the
dimensionless potential in a frame which co-rotates with their mean
motion is reduced to
\begin{equation} 
U = \left(1 - {M_p \over M_{t}}
\right) \left( {1 \over r_1} + {r_1 ^2 \over 2} \right)
+ {M_p \over M_{t}} \left( {1 \over r_2} + {r_2 ^2 \over 2} \right) 
\end{equation}
where $r_1$ and $r_2$ are the dimensionless distances from the host
star and the planet, respectively (\cite{md99}). The
physical values of $r_1$ and $r_2$ are obtained by multiplying them
by $a$,  and that of $U$ can be obtained by multiplying it by $G
M_{t} /a$.

For this potential, there are two locations, L1 and L2, near the
planet which are saddle points.  Both points lie on the line joining 
the planet and the host star, with L1 between them and L2
farther away from the host star than the planet. The distances of
the L1 and L2 points ($D_{L1}$ and $D_{L2}$) to the center of the
planet are
\begin{equation}
D_{L1, L2} \simeq \left[ \alpha \mp {\alpha^2 \over 3} \right] a
\end{equation}
where $\alpha \equiv (M_p/3 M_\ast)^{1/3}$.  Gas that expands beyond the L1
point falls toward and is accreted by the host star,  whereas that
which expands beyond the L2 point spirals outwards to form an excretion
disk.

The equipotential surfaces containing L1 or L2 points are distorted
from spherical symmetry about the planet.  On the planet's ``night
side'' (in the direction of L2), the distance between the planet's
center and the closest point (L1$^\prime$) on the equipotential
surface which contains the  L1 point is
\begin{equation}
D_{L1} ^\prime \simeq D_{L1} - {2 \over 3} \alpha^{3/2} a.
\end{equation}
The distance between the L1$^\prime$ and the L2 points is
\begin{equation}
\Delta D = { D_{L2} - D_{L1} ^\prime} \simeq {2 a \over 3}
(\alpha^{3/2} + \alpha^2 )
\end{equation} 
which for a Jupiter-mass planet is of the order of  a few percent of
$D_{L1}$.  It is possible for the planet's envelope to reach the
equipotential surface containing the L1 point but not that containing
the L2 point.

However, it is also possible for both equipotential surfaces to 
 be contained within the
planet's atmosphere. For an isothermal planetary atmosphere, the
density ($\rho$) scale height 
\begin{equation}
h_\rho = {\rho \over (\partial \rho / \partial r)} = {c_s^2 (r_p)
\over \nabla \Psi}.
\label{eq:hrho}
\end{equation}  
The gradient of the potential near the L1$^\prime$ point is
\begin{equation}
\nabla \Psi = - {G M_{t} \over a^2} {d U \over d r_2} \simeq 4
\alpha^{3/2} {G M_{t} \over a^2}
\label{eq:nablapsi}
\end{equation}
which is somewhat softer than that of isolated planets.  The ratio
\begin{equation}
{h_\rho \over \Delta D } \simeq {9 \over 8} {{\cal R}_g T_p a \over
\mu_g G M_p}
\label{eq:hoverd}
\end{equation}
where $T_p$ and $c_s (r_p)$ are, respectively, the temperature and sound
speed of the planetary surface, and ${\cal R}_g$ and $\mu_g$ are the
gas constant and molecular weight.  On their day side, planets are
heated to an equilibrium temperature $T_p \sim T_\ast (R_\ast / a)^{1/2}$, 
where the surface temperature of solar-type host stars is $T_\ast \sim 6
\times 10^3$ K.  For Jupiter-mass planets at $\sim 0.04$ AU, $h_\rho
\sim \Delta D$, such that both L1 and L2 may be contained in their
atmosphere as the envelope fills the Roche radius.

In the above estimate, we used the day-side equilibrium temperature to
estimate $T_p$.  But both L1$^\prime$ and L2 points are on the
``night-side'' of the planet and are shielded from the stellar
radiation. In principle, the night side of a mature planet could have
a substantially lower surface temperature if  the cooling time scale there
is much shorter than the time scale for sound waves to cross the dark
hemisphere.  But, the resulting longitudinal pressure gradient would
excite large scale circulation with speeds comparable to $c_s$ of the
hot side (\cite{sg02}, Burkert  et al. 2002); it is
not clear yet what the resulting temperature on the dark side would
be.  For planets with relatively young ages ($t_p \approx 10^6 - 10^7$ yr), however, the
effective temperature of unheated Jupiter-mass planets arising from their intrinsic
luminosity (\cite{bur97}) falls in the range 500 -- 800 K.
If the planets migrated too
close to their host stars and became tidally inflated shortly after
they were formed, the stellar heating would introduce little temperature
difference between the day and night sides of the planet.

In addition, if there is sufficiently intense energy dissipation in
some planets to inflate their $R_p$ beyond their $R_L$, their surface
luminosity ${\cal L}_c (R_L)$ would exceed $10^{-5} L_\odot$ (see
\S3.2).  For these tidally inflated planets with $R_p \sim R_L > 2
R_J$, equations  (\ref{eq:equalL}) and (\ref{eq:cale}) are the more
appropriate approximations.  Using a black body scaling law ${\cal L}
/ L_\odot = (R_p/R_\odot)^2 (T_p/T_\odot)^4$ where $T_\odot =5800$K is
the effective temperature of the Sun and equation (\ref{eq:cale}), we find
that when $R_p = R_L$,
\begin{equation}
T_p \simeq T_\odot \left( {{\cal L} \over L_\odot} \right)^{1/4}
\left({ R_L \over 2 R_J} \right)^{\gamma/4 - 1/2} 
\left( {R_\odot \over 2 R_J} \right)^{1/2}
\end{equation}
which for a $M_p = 0.63 M_J$ planet is $\sim 700-1000$ K.  In this
limit, $h_\rho$ may be a fraction of $\Delta D$.  But, we also note
that near both the  L1 and L2 points, the magnitude of $\nabla \Psi$
vanishes so that $h_\rho$ may be substantially larger than that in
equation (\ref{eq:hoverd}) for the  L1$^\prime$ point.

\subsection{Mass flow induced by envelope expansion}
We consider the flow across equipotential surfaces below but near the L1
point.  For computational convenience, we adopt a spherically
symmetric approximation for the equipotential surface.  During the
quasi-static expansion of the envelope, the mass flux advected from
the planet's interior across $U$ is
\begin{equation}
{\dot M}_a (U) \simeq 4 \pi \rho(U) r^2 (U) V_r (U)
\label{eq:mdote}
\end{equation}
where $V_r (U)$ is the outflow speed across the equipotential surface
located at a radius $r (U)$ from the planet's center.  $V_r$ should
not be confused with the size expansion rate $\dot R_p$.  Assuming the
planet's expansion is homogeneous, we can approximate
\begin{equation}
{{\dot M}_a (U) \over M_p} \simeq \lambda {\dot R_p \over R_p} =
{\lambda \over \tau_{\rm R}} 
\label{eq:mdote1}
\end{equation} 
where $\lambda$ is a positive dimensionless structure parameter.  We
show below that for a homogeneous interior, $\lambda=3$ and it is of
the same order but slightly less than 3 for more realistic models.

Neglecting departure from spherical symmetry, we define the
equipotential surface $U$ to be located at a distance  $R_L (t) $ from
the planet's center.  At any given time $t_1$, the mass interior to
$R_L$ is $M_\delta (t_1) = 4 \pi \int_0 ^{R_L (t_1)} \rho(r, t_1) r^2 dr.$
For $R_p > R_L$, the mass outside $R_L$, {\it i.e.} that between $R_L$ and 
$R_p$, would be lost.  At a
later time $t_2 = t_1 + \Delta t$, the planet's density distribution,
size, and Roche radius evolve at rates $\dot \rho$, $\dot R_p$, and
$\dot R_L$ respectively.  The difference in the mass contained in $R_L$
becomes
\begin{equation}
\dot M_\delta (t_1) = {\Delta M_\delta (t_1) \over \Delta t} 
= \left( { M_\delta (t_2) - M_\delta (t_1) \over \Delta t} \right)
\simeq 4 \pi \rho (R_L (t_1)) R_L(t_1)^2 \dot R_L 
+ 4 \pi \int _0 ^{R_L (t_1)} \dot \rho r^2 d r.
\label{eq:a1}
\end{equation}
In a Roche potential,
\begin{equation}
{\dot R_L \over R_L} = {1 \over 3} \left({\dot M_\delta (t_1) \over 4 \pi
\int_0 ^{R_L (t_1)} \rho r^2 d r} \right).
\label{eq:a2}
\end{equation}
From equations (\ref{eq:a1}) and (\ref{eq:a2}), we find
\begin{equation}
\dot M_\delta (t_1) \left( 1 - {\rho (R_1, t_1) R_L(t_1) ^3 \over 3 
\int_0 ^{R_L (t_1)} \rho r^2 d r} \right) =
4 \pi \int _0 ^{R_L (t_1)} \dot \rho r^2 d r.
\label{eq:a3}
\end{equation}
In the limit that $R_p$ is slightly larger than $R_L$, mass loss occurs
near the atmosphere where the density is low and the second term on
the left hand side of equation  ({\ref {eq:a3}) is very small. But, after the
planet's atmosphere is removed, the envelope continues to expand.  The
atmosphere merely responds with an expansion to restore a state of
hydrostatic equilibrium. The actual value for $\rho(R_1, t_1)$ is
that prior to the removal of the envelope between $R_L$ and $R_p$.

An approximate solution can be constructed in which we consider the
limit that $R_L$ is much smaller than the original $R_p$. We first
consider the rate of change of density in terms of the rate of change
of some intermediate reference radius $\dot R_{r}$.  The results of
numerical models of planetary interiors  (see \S5) indicate that the
inflation is nearly homologous, {\it i.e.} the ratio of the half
density radius changes as $R_{1/2} \propto R_p^{\zeta_1}$ where
$\zeta_1 \simeq 0.47$.  The central density $\rho_c \propto
R_p^{\zeta_2}$ where $\zeta_2 \simeq -1.35$.  These numerical results
also indicate that the density within $R_{1/2}$ is nearly homogeneous
such that the mass contained within it $\sim 4 \pi \rho_c R_{1/2}^3/3$
is approximately constant.

In order to derive the rate of change in density $\dot \rho$, we first
consider a reference radius $R_r$ which, at some time $t_1$, contains
a mass $\int_0 ^{R_{r} (t_1)} \rho(t_1) r^2 d r = M_{r}.$ At $t_2 =
t_1 + \Delta t$, the radius which contains the same amount of mass has
evolved to $R_{r} + \dot R_{r} \Delta t$ while the density evolves at
a rate $\dot \rho$ so that $\int_0 ^{R_{r} (t_1) + \dot R_{r} (t_1)
\Delta t} (\rho(t_1) + \dot \rho(t_1) \Delta t) r^2 d r = M_{r}.$ Note
that although $M_p$ may decrease due to mass loss, but, $M_{r}$ is
chosen to attain the same value.  From these expressions of $M_{r}$,
we find
\begin{equation}
\int_0 ^{R_{r} (t_1)} \dot \rho(t_1) r^2 d r
= - \rho (R_{r}, t_1) R_{r} (t_1)^3 {\dot R_{r} (t_1)
\over R_{r} (t_1)}.
\label{eq:b3}
\end{equation}
Note that the rate of change $\dot R_{r}$ includes that due to tidal
dissipation as well as mass loss.  For isolated planets with $M_p >
M_J$, $\partial R_p / \partial M_p < 0$ because the
planet's interior is in a partially degenerate state.  Thus, mass loss
may also lead to size increases.

Without the loss of generality, we can choose $R_{r}$ to be $R_L$ in
the above expression so that we find from equations(\ref{eq:a3}) and (\ref{eq:b3})
that 
\begin{equation}
{\dot M_\delta (t_1) \over M_p } \left( 1 - {4 \pi \rho (R_L) R_L ^3 \over
3 M_p } \right) \simeq {4 \pi \rho(R_L) R_L ^3 \over M_p } \left( {\dot R_r
\over R_r} \right)_{R_L} .
\label{eq:mdelta}
\end{equation}
Setting $\dot M_\delta = \dot M_a$, we find from eqs (\ref{eq:mdelta})
and (\ref{eq:mdote1}) that
\begin{equation}
\lambda \sim {4 \pi \rho (R_L) R_L ^3 \over 
M_p } {(\dot R_r/R_r )_{R_L} \over \dot R_p / R_p}.
\label{eq:lambda1}
\end{equation}
Shortly after overflowing the Roche radius $R_L$, $(\dot
R_r/R_r)_{R_L} \simeq \dot R_p / R_p$ and $\rho (R_L)$ is very small.
But in the advanced stage of Roche lobe overflow, $(\dot R_r/R_r
)_{R_L} \simeq \dot R_{1/2} / R_{1/2} \simeq \zeta_1 \dot R_p / R_p$
but $\rho (R_L)$ may be only slightly smaller than $\rho_c$.  For a
nearly homogeneous planetary interior $\lambda \sim 3$ and if it is
more central condensed, $\lambda < 3$.  Our numerical models for their
interior indicate that under the heating due to the tidal dissipation,
planets establish a quasi hydrostatic equilibrium prior to reaching a
thermal equilibrium.  Since most of the gaseous giant planets'
interior is convective, the temperature gradient is essentially
adiabatic and the average global adiabatic index is $\sim 1$.  In
these models, $\lambda$ is of the same order but slightly less than 3.
The numerical results to be presented in \S5 do not depend sensitively
on the actual value of $\lambda$.

\subsection{Mass flux through the L1 \& L2 points}
As the planet's envelope expands and fills its Roche lobe, gas in its
atmosphere encounters and flows through the saddle-shape nozzle in the
equipotential surface. Near this region, the above approximation based
on spherical symmetry can no longer be applied.  Instead, we consider
a cylindrical geometry with an axis of symmetry along the line
joining the planet and its host star.  The planet's mass loss rate
through the L1 point between the equipotential surface which contains
it and that with an arbitrary value $U_n$ is
\begin{equation}
{\dot M}_{L1} (U_n) \simeq \int_{U_1} ^{U_n} \rho (U) c_s (U) {d \sigma 
\over d U} dU
\label{eq:mdotl1}
\end{equation}
where $\sigma$ is the cross section of the nozzle (Lubow \& Shu 1975,
Pringle 1985).  Note that the total mass flux can be obtained by
setting $U_n = U_\infty$.  In a coordinate system in which $x$ is
defined to be the direction joining the centers of the planet and its
host star, $y$ is orthogonal to $x$ and in the orbital plane, and $z$
is orthogonal to both $x$ and $y$, the narrowest opening of the nozzle
occurs at $x \simeq x_{L1}$, with $\sigma = \pi (s a)^2$ lies in the
$y-z$ plane perpendicular to the $x$ axis and $s a$ is the radius of
the nozzle. The dimensionless variable $s$ is a function of U and it
vanishes for $U( r < D_{L1})$.  In the neighborhood of L1, $\partial
U/\partial s \simeq 3 s$ so that
\begin{equation}
{d \sigma \over d U} = {2 \pi a^2 s \over(d U / ds)} \simeq {2 \pi a^2
\over 3}.  
\end{equation}
For the same region, we find from equation (\ref{eq:hrho}) that 
\begin{equation}
\rho(s) = \rho_1 {\rm exp} \left( - {3 G M_t s^2 \over 2 a c_s^2} \right)
\end{equation}
where $\rho_1$ is the density at the L1 surface, and from 
equation (\ref{eq:mdotl1}), we find
\begin{equation}
\dot M_{L 1} (s_n) = {2 \pi \rho_1  a^3 c_s ^3 \over 3 G M_t} 
\left[ 1 - {\rm exp} \left( - {3 G M_t s_n^2 \over 2 a c_s ^2 } \right)
\right]
\end{equation}
where $s_n= [2 (U_n - U_1)/3]^{1/2}$.

The expansion of the planet's envelope would continue so long as  ${\dot
M}_a (U) > {\dot M}_{L1}$.  As the planet's envelope climbs out of the
potential well, $\rho$ would increase which leads to enhanced spillage
via the L1 point.  We now estimate the critical condition which may
lead to Roche lobe overflow via the L2 point.  The difference in the
potential between L1 and L2 points is
\begin{equation}
\Delta U \simeq 2 \alpha^3
\label{eq:deltau}
\end{equation}
so that $s_{L2} = 2 (M_p/M_\ast)^{1/2}/3$ and 
\begin{equation}
\dot M_{L 1} (s_{L2}) = {2 \pi \rho_1  a^3 c_s ^3 \over 3 G M_t} 
\left[ 1 - {\rm exp} \left( - {2 G M_p \over 3 a c_s ^2 } \right)
\right]
\label{eq:mdl1}
\end{equation}
The expansion of the envelope would terminate if a mass transfer
equilibrium can be established in which
\begin{equation}
{\dot M}_{L1} = - {\dot M}_a (U(L1))
\label{eq:equim}
\end{equation}
if we neglect
the adjustment of the equipotential surface due to mass loss and
orbital evolution (for further discussions, see next subsection).

From equations  (\ref{eq:mdote1}), (\ref{eq:mdl1}), and (\ref{eq:equim}), we
find that the mass transfer equilibrium via the L1 point would only
attainable if the local density
\begin{equation}
\rho (L1) \gtrsim \rho_c = \left({ 2 \lambda \over 3} \right) \left( {\dot
R_p \over c_s (L1)} \right) \left( { G M_p \over R_p c_s^2} \right) \rho_a
\label{eq:rhol1}
\end{equation}
where the planet's average density within its Roche lobe 
$\rho_a = 3 M_p / 4 \pi R_L^3
\simeq 10^{-2}$g cm$^{-3}$ for a Jupiter mass planet at 0.04 AU.
Since the envelope of the inflated planet expands  at a speed $\dot R_p
< < c_s$, $\rho(L1) < < \rho_a$.  For $R_p = R_L$ and $e \sim 0.14$,
equations(\ref{eq:taur}) and (\ref{eq:rhol1}) lead to
\begin{equation}
\rho_c \sim {7 \lambda \beta e^2 \over 6 q_p Q_p ^\prime} \left({ G M_p \over 
c_s ^2 R_L } \right) \left( {R_L n \over c_s } \right)
\rho_a \sim 10^{-6} \rho_a \sim 10^{-8} {\rm g} \ {\rm cm}^3.
\label{eq:rhol1a}
\end{equation}
Since $\rho_a \propto a^{-3}$, $c_s \propto a^{-1/4}$, and $n \propto
a^{-3/2}$, $\rho_c$ is proportional to $a^{-15/4} e^2$ and it would exceed
$10^{-7}$g cm$^{-1}$ at $a < 0.025$AU, which corresponds to a  1.5 day period
around a 1 $M_\odot$ G dwarf.

When planets overfill their Roche radius, $\rho(L1)$ is comparable to
the density $\rho_{atm}$ at their effective radius where the optical
depth is of the order unity.  Although $\nabla \Psi=0$ at the L1
point, it is $\sim G M_p / R_L ^2$ or that given in
equation (\ref{eq:nablapsi}) elsewhere on the equipotential surface which
contains that point.  At $a=0.04$ AU where the day-side temperature of
$T_p \sim 1-2 \times 10^3$K, we find from equation (\ref{eq:hrho}) that
$h_\rho \sim 2-3 \times 10^9$ cm for a Jupiter-mass planet.  For this
$T_p$, the opacity of the atmospheric gas is $\kappa \sim 10^{-1}$
cm$^2$ g$^{-1}$ so that $\rho_{atm} \sim 10^{-8}$g cm$^{-3}$.  For
modest $e$ values and $a > 0.04$ AU, $\rho_{atm} > \rho_c$ such that
the optically thin atmosphere engulfs both the L1 and L2 points.
In this limit, the density at the L2 point can be derived from equations
(\ref{eq:hrho}) and (\ref{eq:deltau}) to be
\begin{equation}
\rho_2 = \rho_1 {\rm exp} \left( - {2 G M_p \over 3 a c_s^2} \right).
\end{equation}
On a $M_p=1 M_J$ planet at 0.04 AU from a G dwarf where $T_s \sim 10^3
K$, $\rho_2 \sim 0.2 \rho_1$ so that gas near the Roche lobe mostly
flows  toward the host star via the L1 point.

From equation (\ref{eq:rhol1a}) we find $\rho_c > \rho_{atm}$ for planets with
\begin{equation}
a < a_{L2} = \left[ \left({3 \rho_\ast \over \rho_{atm}} \right)
\left( {G M_p \over c_\ast^2 R_\ast} \right)
\left( {7 \lambda \beta \over 6 q_p Q_\ast ^\prime} \right)
\left( {R_\ast n_\ast \over c_\ast} \right) e^2 \right]^{4 / 15} R_\ast 
\label{eq:al2}
\end{equation}
where $\rho_\ast = 3 M_\ast / 4 \pi R_\ast^3$, $n_\ast=(GM_\ast
/R_\ast^3)^{1/2}$, and $c_\ast = ({\cal R}_g T_\ast/ \mu_g)^{1/2}$.
For $e \sim e_R \sim 0.14$, $a_{L2} \simeq 0.03 AU$.  Equivalently,
with a given $a$, $\rho_c > \rho_{atm}$ for planets with
\begin{equation}
e > e_{L2} = \left[ \left({\rho_{atm}
\over 3 \rho_\ast} \right) \left( {c_\ast^2 R_\ast \over G M_p}
\right) \left( {6 q_p Q_\ast ^\prime \over 7 \lambda \beta} \right)
\left( {c_\ast \over R_\ast n_\ast} \right) \left({ a \over R_\ast}
\right) ^{15/4} \right]^{1 / 2}.
\label{eq:el2}
\end{equation}
At $a=0.03$ AU, $e_{L2} \simeq 0.14$.  In this limit, the opaque
regions of the envelope extend above the equipotential surface which
contains the L1 point. The temperature and density scale height in the
envelope are greater than their surface values so that the density
differential between the L1 and L2 points is reduced.  Consequently,
gas at the planets' effective radius may flow through both L1 or L2
points with comparable rates.  An accurate determination of the ratio
of the mass flux and the specific angular momentum flux through the 
L1 and L2 points needs to be analyzed
with more detailed numerical hydrodynamic simulations (Burkert \& Lin
2003).

In \S3, we showed that the necessary and sufficient criteria for
planetary inflation are $e>e_r$ and $e>e_L$. The instantaneous tidal
radius of eccentric planets may modulate by a fraction $e$.  At the
periastron, gas outside the planet's minimum tidal radius is removed.
During other phases, the planet's atmosphere is refilled as it adjusts,
on its internal dynamical time scale, to an evolving quasi
hydrostatic equilibrium,  while the underlying envelope re-expands  on a
much longer internal heating time scale.  If the ratio
\begin{equation}
{h_\rho \over 2e D_{L1} } \simeq
{(3 \alpha)^{1/2}\over 4 e} {{\cal R}_g T_p a \over \mu_g G M_p}
\end{equation}
is comparable to or larger than unity, the planet's atmosphere would
overflow the Roche lobe at all orbital phases and the mass loss
would be continuous.  Otherwise, bursts of mass loss would occur
during each periastron passage.

\subsection{The rate of planets' mass overflow through their Roche lobe}
Planets' Roche lobe overflow also leads to their orbital migration.
If the overflow occurs through both L1 and L2 points, angular momentum
exchange between the infalling and outflowing transferred gas would
result in orbital migration. In general,
mass loss would lead to a reduction in the Roche radius at a rate
\begin{equation}
{\dot R_L \over R_L} = \left({({\dot M}_{L1} + {\dot M}_{L2}) 
\over 3 M_p} + {\dot a \over a} - {\dot e \over 1-e} \right)
\label{eq:dotrl}
\end{equation}
along with the equipotential surfaces which contain the L1 and L2
points.  The shrinkage of the Roche lobe would enhance mass flux across
the equipotential surfaces which contain both the L1 and L2 points by
an amount
\begin{equation}
{ {\dot M}_R \over M_p} \simeq - \lambda {\dot R_L \over R_L}  
\label{eq:mdotep}
\end{equation} 
such that the effective mass flux across the equipotential surface of the
newly adjusted Roche lobe $U^\prime (L1)$ is 
\begin{equation}
{\dot M}_e (U^\prime (L1)) = {\dot M}_a (U(L1)) + {\dot M}_R.
\label{eq:mdtepu}
\end{equation}
In a mass transfer equilibrium, this mass flux is lost through the L1 and
L2 points such that
\begin{equation}
({\dot M}_{L1} + {\dot M}_{L2}) = -{\dot M}_e (U^\prime (L1)).
\label{eq:medd}
\end{equation}

For $e > e_{L2}$, mass is lost via both the L1 and L2 points. In
this limit with the simplified
assumption $\dot a \simeq 0$ \footnote{This over-simplified
assumption is just for the demonstration of this possible scenario,
and will be used in the next section. What exactly happens to $a$ in 
this limit is undetermined in this paper as stated in the 
preceding section}, we find, from equations(\ref{eq:mdote1}),
(\ref{eq:dotrl}), (\ref{eq:mdotep}), (\ref{eq:mdtepu}), and
(\ref{eq:medd}), the planets' mass loss rate to be
\begin{equation}
{\dot M}_{L2} = - \left( { \lambda M_p \over 1 - \lambda/3} \right)
\left({\dot R_p \over R_p} + {\dot e \over (1-e)} \right).
\label{eq:dotr2}
\end{equation}
For planets with a relatively flat density distribution ({\it i.e.} 
$\lambda$ is smaller but close to 3), the magnitude of ${\dot M}_{L1}$
can greatly exceed that of ${\dot M}_a (U(L1))$.

The reduction of the planet's mass reduces the radial velocity
amplitude $K$ for Earth-based observers which is determined by
\begin{eqnarray}
K &=&{M_p \sin i\over M_{t}}\left(
{2\pi G (M_{t}) \over P} \right)^{1/3}
(1-e^2)^{-1/2} \nonumber \\
&\approx& 1.41 
\left( { M_p \sin i \over 0.01 M_J } \right)
\left( { M_\odot    \over M_\ast }   \right)^{2/3}
\left( { 3\,{\rm days} \over P }     \right)^{1/3} 
(1-e^2)^{-1/2}\,{\rm m s^{-1}},
\end{eqnarray}
where $P$ is the planet's period and $i$ is the inclination angle of
the orbit with respect to the line of sight. For an observational
limit of 1 m s$^{-1}$, planets with a few Earth masses  and $P\lesssim 3$ days 
would be undetectable with the Doppler method.  Thus, a substantial
mass loss may account for the absence of any extrasolar planets with
$a \lesssim 0.04$AU.

If, however, $e < e_{L2}$, planets would overflow their Roche lobes 
mostly via the L1 point and they would move outwards as they absorb
the angular momentum and energy from the transferred gas which flows
toward their host stars. This dynamical property is analogous to
conservative mass transfer from a low-mass to a high-mass star in an
interacting binary system.  Assuming all the transferred material is
accreted by the host stars and the total angular momentum of the
system is conserved, the semi-major axes of these planets would expand
at a rate
\begin{equation}
{ {\dot a}_m \over a} = 2 {e {\dot e}_m \over 1 - e^2} 
-2 { {\dot M}_{L1} \over M_p} .
\end{equation}
Unless the planets have very small $a$'s, the matter lost from their
Roche lobe cannot directly strike the surface of the host star, and
it forms a circumstellar disk.  During its infall onto and subsequent
viscous evolution within the disk, both energy and angular momentum
are tidally transferred between the overflow gas and the planets'
orbits.  Unless the net energy and angular
momentum transfer rates are exactly balanced, eccentricity of the
planet may be modified at a rate $ {{\dot e}_m / e} = \eta { {\dot
M}_{L1} / M_p}.$ The actual magnitude and sign of $\eta$ is determined
by the combined and competing effects of corotation versus Lindblad
resonances (Goldreich \& Tremaine 1980).  Despite the mass transfer,
the binary stars' orbits in cataclysmic variables remain circular,
which implies that the magnitude of ${\dot e}_m$ is negligibly small
compared with that due to the tidal circularization in
equation (\ref{eq:edot1}).  For the remaining discussions, we neglect the
contribution associated with ${\dot e}_m$.

The semi-major axis of an eccentric orbit would shrink due to tidal
orbital circularization at a rate $\dot a_e=2ae {\dot e}/(1-e^2)$ (see
eq. \ref{eq:taua}). This effect of circularization on $a$ is more
prominent for larger $e$, which would be the case for extrasolar
planets.  The transfer of angular momentum between the planet's orbit
and the stellar spin also introduces $\dot a_e= 2 a \dot J_\ast/J_o$
(see eq.  \ref{eq:adot}) which is generally much smaller than that
induced by mass loss or orbital circularization.  Finally, the
planets' tidal interaction with their residual nascent disk also
contributes to inward migration on a time scale $\tau_d$.
Consequently, the total change of $a$ comes from these three effects:
\begin{equation}
{\dot a \over a}={\dot a_m +\dot a_e \over a} =-2{\dot M_{L1} \over
M_p} +2 {e {\dot e} \over (1-e^2)} + 2 {\dot J_\ast \over J_o} - 
{1 \over \tau_d}.
\label{eq:dota}
\end{equation}
The above equation, together with equation (\ref{eq:dotrl}), implies that
the Roche lobe changes at a rate
\begin{equation}
{{\dot R_L} \over R_L} = - {5 {\dot M}_{L1} \over 3 M_p}
-{{\dot e}  \over 1+e}
\label{eq:dotr2b}
\end{equation}
where we have neglected the effects of stellar spin synchronization
and planet-disk tidal interaction.  The first term on the right hand
side of the above rate equation indicates that the retreat of the
planet's orbit due to mass overflow would enlarge the Roche lobe.  On
the other hand, the shrinkage of the planet's orbit due to orbital
circularization would reduce the Roche lobe in size as is indicated by
the second term.

In a mass transfer equilibrium, we find, from equations (\ref{eq:mdote1}),
(\ref{eq:mdotep}), (\ref{eq:mdtepu}), (\ref{eq:medd}), and
(\ref{eq:dotr2b}), the mass transfer rate via the L1 point to be
\begin{equation}
{\dot M}_{L1} = - { \lambda M_p \over (1 + 5 \lambda/3)}  
\left( {\dot R_p \over R_p} + { {\dot e}\over 1+e}
\right).
\label{eq:dotr2c}
\end{equation}
Since $\lambda$ is positive and greater than unity, the magnitude of
${\dot M}_{L1}$ is significantly reduced from that of ${\dot M}_a
(U(L1))$.

\section{Mass and dynamical evolution due to Roche lobe overflow}
The discussions in the two previous sections indicate that the
evolution of $M_p$, $e$, and $a$ are closely connected to each other.
In this section, we present analytic approximations  to assess the
dynamical evolution and the outcome of planets which undergo tidal
inflation instability. 

\subsection{Five stages of planets' tidal inflation}
For illustration purposes, we divide the evolution into five stages.

{{\bf Stage 1:} \it Runaway inflation.}
We consider a planet with an initial semi major axis $a_i$,
eccentricity $e_i$, and radius $R_i$, so that its initial angular
momentum is $J_i = M_p (G M_\ast (1 - e_i^2) a_i)^{1/2}$.  These
parameters are determined by processes which brought the planets to
the vicinity of their host stars.  For example, during the epoch of
planet formation, young solar type stars may have a radius 2-3
$R_\odot$ so that the planet-star and planet-disk interaction would
lead to a halting radius in equation (\ref{eq:ahalt}) $a_i = R_{\rm halt} 
\sim 0.01-0.05$ AU. Or if planets are scattered to the vicinity of
their host stars, their $a_i$ and $e_i$ may be relatively large
but their periastron distance may be only a few stellar radii.
 
Provided $R_i < R_c \simeq 2 R_J$, the necessary and sufficient
conditions for tidal inflation instability are, $e_i > e_R(R_c) > e_L
(R_c)$ for $a_i < a_{RL} (R_c)$ (see eq. \ref{eq:aRL}) and $e_i > e_L
(R_c) > e_R (R_c)$ for $a_i > a_{RL}$.  For $R_i > R_c$, these
conditions are $e_i > e_R(R_i) > e_L (R_i)$ for $a_i < a_{RL} (R_i)$
and $e_i > e_L (R_i) > e_R (R_i)$ for $a_i > a_{RL} (R_i)$.
Unstable planets with $a_i < a_c$ overflow their Roche lobe prior to
runaway inflation whereas those with $a_i > a_c$ undergo runaway
inflation to overflow their Roche lobe.  

These criteria for runaway tidal inflation are  derived under the
assumption that $e$ is not damped significantly as the planets adjust
their sizes toward $R_c$.  This approximation is generally adequate
for $e>> e_R$.  But if $e$ is marginally larger than $e_R$ initially,
it may be damped by an non negligible amount and become stabilized
before $R_i$ can reach $R_c$.  Such  a possibility is best analyzed
numerically (see \S6).

The dimensionless expansion rate
\begin{equation}
{\dot R_p \over R_p} \approx \left( {e_L^2 - e^2 
\over e_R^2} \right) \left( { \dot e \over e} \right) 
\label{eq:ttaurr}
\end{equation}
is obtained from equation (\ref{rd}) by interpreting $\dot R_p$ as the
adjustment rate of the planets' half-mass rather than the Roche
radius. Note that strictly speaking, $e_L$ in the above equation should
not be the value related to ${\cal L}_e$ as shown in eq(\ref{eq:eL}) but
the value related to $\cal L$. However, our internal structure code indicates
that ${\cal L} \simeq {\cal L}_e$ for a given $R_p$ and $M_p$, and
we adopt this assumption hereafter in this section.
Equation (\ref{eq:ttaurr}) is appropriate for all $R_p$ and $a$'s.
For $e > e_R$, the planets' radius adjustment time scale, $\tau_R =
R_p / \vert \dot R_p \vert < \tau_e$ at the onset of tidal inflation.
Since $\tau_R$ is a rapidly decreasing function of $R_p$ and the planets'
semi major axes evolve only after they overflow their Roche radius, 
the duration of stage 1 is determined by $\tau_R$ with $R_p = R_i$
such that 
\begin{equation}
\tau_1 \simeq \tau_R (a_i, R_i) = {R_i \over \dot R_p (a_i, R_i)}
={10} \left( {e_R^2 \over e_L^2 - e^2 } \right)
\left( {Q_p^\prime \over 10^6} \right)
\left( {M_p\over M_J} \right) \left( {\msun \over M_\ast} \right)^{3/2}
\left( {a \over 0.04\,{\rm AU}} \right)^{13/2}
\left( {2 R_J \over R_i} \right)^5
\, {\rm Myrs}.
\label{eq:tau11}
\end{equation} 
which, for $R_i \sim R_c \sim 2 R_J$, is longer than the expected
values of $\tau_d (\sim 10^{5-6}$ yrs).  Thus, it would be a reasonable
approximation to neglect the effect of planets' tidal interaction with
the disk if they have already terminated their accretion and
contracted to $< 2 R_J$ when they migrated to the proximity of their
host stars. However, if the protoplanets' orbital migration, induced
by their interaction with the disk occurs on a time scale which is
comparable to or shorter than that for them to to contract (Trilling
{\it et al.} 1998), they may reach the vicinity of their host stars
with $R_p \sim 2R_J$.  

Equation (\ref{eq:tau11}) indicates that stage 1 could be completed for those
planets which migrated to $a<0.04$ AU while their host stars evolved 
through the classical and weak-line T Tauri phases.  At the end of
stage 1, the tidally inflated planets lose mass through Roche lobe
overflow (see stage 2 below).  All the metal-rich planetary tidal
debris flowing out of the L1 point is accreted by the host stars and
thoroughly homogenized in their surface convective envelopes.  Early
planetary disruption and stellar accretion is unlikely to
significantly modify the metallicity of the host stars' convection
zone because it is extensive.  But $\tau_1 > 10^7$ yrs for planets
which terminated their migration at slightly larger radii (at $a>
0.05-0.06$AU).  In this limit, the onset of the planets' Roche lobe
overflow is delayed until their host stars have fully evolved onto the
main sequence.  For example, the total mass of the surface
convection zone of a G dwarf host star is reduced to an asymptotic
value of $\sim 0.02 M_\odot$ after $\sim 30$ Myr (Ford {\it et al.}
1999). Thereafter, the accretion of the metal-rich tidal debris from
the Roche-lobe overflowing Jovian-mass planets may significantly
contaminate the outer regions of their host stars.  Most planet-bearing
stars are observed to have metallicity higher than the field stars
(Gonzales \& Laws 2000). 

{{\bf Stage 2:} \it Orbital evolution of Roche-lobe filling planets}
($e^2 - e_L ^2 (R_L) > e_R ^2 (R_L)$).
After filling its Roche lobe,  an inflated planet continues to
expand and to lose mass through both L1 and L2 points at a rate
\begin{equation}
{\dot M_{L2} \over M_p} = - {\lambda \over (1 - \lambda /3)} \left( 
{\dot R \over R_p} + {\dot e \over 1 - e} \right) .
\label{eq:moreel2}
\end{equation}
with $\dot a \simeq 0$ if their $a_i < a_{L2}$ (see eq. 
\ref{eq:dotr2}). But, for $a_i > a_{L2}$ the mass loss and migration
rates are
\begin{equation}
{\dot M_{L1} \over M_p} = -{\lambda \over (1 + 5 \lambda /3)} \left( 
{\dot R_p \over R_p} + {\dot e \over 1 + e} \right) 
\label{eq:lessel1}
\end{equation}
\begin{equation}
{\dot a \over a} =  2 { e {\dot e} \over (1 - e^2)} 
-2 {\dot M_{L1} \over M_p}
\label{eq:lessel2}
\end{equation}
(see eqs. \ref{eq:dotr2c} \& \ref{eq:dota}).
Note that negative values  of $\dot M_{L1}$ correspond to outflow.
Since accretion cannot occur in this context, $\dot M_{L1}$ must be
set to zero if the right hand side of equation (\ref{eq:moreel2}) or
(\ref{eq:lessel1}) becomes  positive.

Neglecting  the effect of $e$ excitation due to planets' interaction
with the disk and the dissipation within the stellar interior, the rate
of $e$ damping is 
\begin{equation}
{\dot e \over e} = -{21 \pi \over 3^{2/3} 2 Q_p ^\prime } \left({M_p
\over M_\ast } \right)^{2/3} { 1 \over P} 
\label{eq:eovere}
\end{equation}
for both limits (see eq. \ref{eq:edot1}).  The dimensionless expansion
rate in equation (\ref{eq:ttaurr}) implies that, regardless of the mass loss,
the expansion of the inflated planets continues provided their $R_e >
R_L$ (negative values of $\dot R$ correspond to contraction.)  Note
that for $e > e_R$, the planets' radius adjustment time scale, $\tau_R
= R_p /\vert \dot R_p \vert < \tau_e$ at the onset of tidal
inflation.  Since $e_R \propto R_p^{-1/2}$, these inequalities are
preserved as the planets inflate and fill their Roche radius.  For
planets with $R_p = R_L$,
\begin{equation}
e_R (R_L) \simeq 0.1 \left( {q_p/ \beta \over 0.75} \right)^{1/2}
\left( {M_p \over M_J} {M_\odot \over M_\ast } \right)^{1/3}
\label{eq:eRRL}
\end{equation}
which is roughly $\propto a^{-1/6}$ according to
eq(\ref{eq:lessel2}) and therefore is weakly dependent on $a$. 
On the other hand,
\begin{eqnarray}
e_L(R_s = R_L) & = & {0.056 \over 26^{(5-\gamma)/6}} 
\left({Q_p ^\prime \over 10^6}{{\cal L}_o \over 10^{-6} L_\odot} \right)^{1/2}
\left[ {M_\odot \over M_\ast} \left({ a \over 0.04 {\rm AU} } \right)^3
\right]^{(5 + 2 \gamma)/12} \nonumber \\
& \quad & \simeq {27.4^{\gamma/6} \over 263.9}
\left({Q_p ^\prime \over 10^6}{{\cal L}_o \over 10^{-6} L_\odot} \right)^{1/2}
\left( {P \over 3 {\rm days} } \right)^{(5 + 2 \gamma) /6}
\label{eq:elrL}
\end{eqnarray}
where the rough approximation is made that the planet is in thermal
equilibrium.  For a $M_p = 0.63 M_J$ planet (with $\gamma=3$, ${\cal
L}_o = 10^{-6} L_\odot$, and $R_p = R_L$),
\begin{equation}
e_L (a, R_L) = 0.02 \left( {Q_p ^\prime  \over 10^6} 
\right)^{1/2} \left( {P \over 3 {\rm days}} \right)^{11/6}.
\label{eq:eLRL}
\end{equation}
which is $\propto a^{11/4}$.  

During stage 2, $e^2 - e_L^2 (R_L) > e_R^2$ so that $\dot R_p$ and
$\dot e$ have opposite signs, {\it i.e.} eccentricity damping leads to
further inflation.  Consequently, the planets are inflated to fill
their Roche radius.  In addition, the magnitude of $\tau_R$ is smaller
than that of $\tau_e$, {\it i.e.} the planets' size adjusts faster
than their eccentricity changes.  For large $e_i$ ($ > > e_R (a_i)$
and $> > e_L(a_i)$), equation (\ref{eq:ttaurr}) implies negligible changes in
$e$ from $e_i$ while a sizable fraction of $M_p$ is lost and $a$ is
increased significantly.  But for $e_i$ marginally larger than
$e_R(a_i)$ and $e_L(a_i)$, the planets' tidal inflation and mass loss
is accompanied  by an  $e$ reduction by a comparable fraction.

Note that if mass loss does not significantly modify $a$, as is the
case for $a_i < a_{L2}$, $e_L$ would remain constant. Consequently,
the stage of runaway inflation continues until the planets are
disrupted locally. But for $a_i > a_{L2}$, most of the mass is lost
through the L1 point while both $a$ and $e_L(R_L)$ increase.

At the end of stage 2, $e^2 (a_L) = e_R ^2 (R_L) + e_L ^2 (a_L, R_L)$
which occurs at $a = a_L$ (see expression for $a_L$ and the
corresponding period $P_L$ in the discussion for stage 3 below).
While mass transfer is sustained, equations (\ref{eq:ttaurr}),
(\ref{eq:lessel1}), and (\ref{eq:lessel2}) implies that
\begin{equation}
{\dot a \over a} = \left( {2 e^2 \over 1 - e^2} + {\lambda \over 
( 1 + 5 \lambda / 3)} \left( {e_L ^2 - e^2 \over e_R ^2 } + {e 
\over 1 + e} \right) \right) {\dot e \over e} \sim {\lambda \over 
( 1 + 5 \lambda / 3)} \left( {e_L ^2 - e^2 \over e_R ^2 } \right)
{\dot e \over e}
\end{equation}
\begin{equation}
{\dot M_L \over M_p} = - {\lambda \over 
( 1 + 5 \lambda / 3)}\left( {e_L ^2 - e^2 \over e_R ^2 } + {e 
\over {1 + e}} \right)  {\dot e \over e}
\end{equation}
Since $\lambda \sim 3$, $\tau_M$ and $\tau_a$ are $<\tau_R$, but 
comparable to $\tau_e$ which is an increasing function of $a$ and $P$.
Thus, the duration of stage 2 is
\begin{equation}
\tau_2 \sim \tau_e (a_L, R_L)
= 0.44 e^{6/11} \left({Q_p
^\prime \over 10^6} \right)^{8/11} \left( {M_\ast \over M_\odot}
\right)^{2/3} \left({M_J \over M_p} \right)^{2/3} {\rm Myr}. 
\label{eq:tealrl}
\end{equation}
which is much shorter than the duration of stage 1, {\it i.e.} $\tau_1 
\simeq \tau_R (a_i, R_i)$.  

Thus, runaway tidally inflation requires a substantial lead-in time
which is followed by a brief expansion and orbital migration
phase.  Because they are substantially inflated, the theoretical
probability that the transit of Roche-lobe overflowing planets in
front of their host stars are observable by distant observers is
\begin{equation}
P_{t 2} \simeq {\pi \over 2} {R_\ast+ R_L (a) \over a} > \left( 
{M_p \over 3 M_\ast}  \right)^{1/3}
\end{equation}
which is larger than  for the present day short-period planets due to the
enlarged sizes of both $R_\ast$ (for $\tau_1 < 10$ Myr) and $R_p$.
The actual probability of observing such events may be much smaller
because 1) only a small fraction of stars have short-period planets
and 2) the duration of this Roche-lobe overflowing stage 2 is brief
compared with the life span of the stars. Note that during the
transit, the host stars' magnitude is reduced by $\Delta m_t \sim 5
{\rm log}_{10} ( {R_p / R_\ast})$ which for the Roche-lobe overflowing
planets may be substantially larger than the present-day short-period
planets.

{{\bf Stage 3:} \it Stagnation and eccentricity damping} ($e_R ^2
(R_L) > e^2 - e_L ^2 (R_L) > 0$). The
discussion below on stages 3-5 is relevant only for the cases with
$a_i>a_{L2}$.  Although the magnitude of the  $e$ reduction depends on the
initial conditions, $e_L ^2 (a, R_L)$ increases monotonically to $e^2 -
e_R ^2 (R_L)$ such that 
\begin{equation}
{e_i ^2 \over e_L ^2 (a_i, R_c) } \left( {e^2 (a_L) - e_R ^2 (R_L) \over e_i^2}
\right) = {e_L ^2 (a_L, R_L) \over e_L ^2 (a_i, R_c)}
\simeq \left( {a_L \over a_i} \right)^{15/2} \left({R_c \over R_L (a_L) } 
\right)^{5-\gamma}.
\label{eq:aiaf}
\end{equation}
In deriving the above equation, we ignore the slight dependence
of ${\cal L}_0$ on $M_p$ for the simple illustration.
Note that $e_R (R_L)$ weakly depends on $a$ (see eq
\ref{eq:eRRL}) and it is less than $e_R (R_c)$ (see eq. \ref{eq:er}).
The onset of tidal inflation instability also requires $e_i > e_L (a_i,
R_c)$ so that the left hand side of equation (\ref{eq:aiaf}) is greater than
unity.  For a $M_p = 0.63 M_J$ planet ($\gamma=3$),
\begin{equation}
a_L = 0.17 \left(e^2 (a_L) - e_R ^2 (R_L) \right)^{2 / 11} 
\left( {M_\ast \over 
M_\odot } \right)^{1/3} {\rm AU}
\label{eq:aL17}
\end{equation}
with a corresponding period
\begin{equation}
P_L = 25.5 \left( e^2 (a_L)- e_R ^2 (R_L) \right)^{3/11} 
\left({10^6 \over Q_p 
^\prime} \right)^{3/11} {\rm days}.
\label{eq:pL25}
\end{equation}

In the limit $e_i > > e_R (R_L)$, the planet's $e$ retains its initial
magnitude $e_i$ when the above conditions are established so that
$a_L \simeq 0.17 e_i ^{4/11} (M_\ast/M_\odot)^{1/3}$AU, and $P_L 
= 25.5 e_i ^{6/11} (10^6/Q_p ^\prime)^{3/11}$ days. But, for $e_i$
marginally larger than $e_R(R_c)$ and $e_L (R_c)$, $a_L \sim a_i$,
{\it i.e.} the planets would not migrate significantly before its 
tidal inflation is stalled.  

As $e^2 - e_L^2 (R_L) - e_R ^2 (R_L)$ reduces to  just below zero,
$\tau_R$ becomes long compared with $\tau_e$, although the planet
continues  to expand provided $e > e_L (R_L)$.  But, the reduction of
$e$ also leads to a reduction in $a$ and $e_L$.  The planet would
continue to fill its Roche lobe and retreat in $a$ if $\partial e_L
(R_L) /\partial e > 1$.  From equation (\ref{eq:eLRL}), we find
\begin{equation}
{\partial e_L (R_L) \over \partial e} = 
{\partial e_L (R_L) \over \partial a} {\partial \dot a \over \partial 
\dot e}
\end{equation}
But, from equations(\ref{eq:lessel1}), (\ref{eq:lessel2}), and
(\ref{eq:ttaurr}), we find
\begin{equation}
\left( {\dot a \over a} \right) = 2 \left( {e^2 \over 1 - e^2} +
\left( {\lambda \over 1 + 5 \lambda/3} \right) \left(  {e_L^2 - e^2
\over e_R^2} + {e \over 1 + e} \right) \right) {\dot e \over e}
\end{equation}
so that with $e^2 = e_R ^2 + e_L ^2$, 
\begin{equation}
{\partial e_L (R_L) \over \partial e} = {11 \over 2} \left( {e^2
\over 1 - e^2} - {\lambda \over (1 + 5 \lambda/3)}  {
1 \over (1 + e)} \right)
\end{equation}
which, for $e<0.6$, is less than unity, primarily due to the planet's
retreat in $a$ as a consequence of angular momentum conservation.
Thus, for modest values of $e$, the quasi-equilibrium cannot be
maintained and $e$ rapidly declines below $e_L$.

During stage 3 when the planet's $e$ decreases from $(e_R^2 +
e_L^2)^{1/2}$ to $e_L$, $\dot R_p$ remains positive but $\tau_R$  is larger
than $\tau_e$. For
\begin{equation}
a < a_{RL} \simeq 0.08 \left( {q_p/ \beta \over 0.75} {10^6 \over Q_p
^\prime} \right)^{2/11} \left( {M_p \over M_J} \right)^{8/33} \left(
{M_\ast \over M_\odot} \right)^{7/33} {\rm AU},
\end{equation}
$e_R > e_L$ so that $e$ decreases by a substantial fraction ($\sim
e_R/e_L$), $M_p$ decreases and $a$ increases by a modest amount.  But
for $a > a_{RL}$, $e_R < e_L$ so that $e$, $M_p$, and $a$ do not
change significantly during stage 3.  The duration of this stage is 
$\tau_3 \sim \tau_e (a_L, R_L)$, {\it i.e.} comparable to $\tau_2$.

From equations (\ref{eq:lessel1}) and (\ref{eq:ttaurr}) we find that Roche
lobe overflow is quenched when
\begin{equation} 
e_L^2 (R_L) -e^2 +{e \over 1+e} e_R(R_L)^2 =0.
\end{equation}
For relatively small $e_R (R_L)$, the solution for the above algebraic
equation is 
\begin{equation}
e = e_t \simeq e_L (R_L) \left( 1 + {e_L (R_L) \over 
2 (1 + e_L (R_L)) } \left[ {e_R (R_L) \over e_L (R_L)} \right]^2  \right).  
\end{equation}
The termination of Roche-lobe overflow is the result of two competing
processes.  The reduction in $e$ leads to decreases in $a$ but an
increase in $R_L$.  For $e>e_t$, the expansion of $R_p$ is faster than
$R_L$, and equation (\ref{eq:lessel2}) corresponds to the conservation of
total angular momentum.  Thus,
\begin{equation}
M_p (a_t, e_t) = M_i \left({ a_i (1 - e_i^2) \over a_t (1 - e_t ^2)
} \right)^{1/2}
\label{eq:mpatet}
\end{equation}
which is the asymptotic mass of the planet.  For $e$ slightly less
than $e_t$, the expansion of $R_L$ is faster than $R_p$.  Thereafter,
equation (\ref{eq:lessel1}) is replaced by $\dot M_{L1} =0$ and
equation (\ref{eq:lessel2}) is reduced to a requirement for subsequent $a$
and $e$ to obey the conservation of specific angular momentum, {\it
i.e.} $a ( 1 - e^2)= a_t (1 - e_t^2)$ where $a_t$ is the semi-major
axis when $e$ is reduced to $e_t$. Because the $e$ range in stage 3 
is limited and $a_L$ has a weak $e$ dependence, $a_t \sim a_L$.

{{\bf Stage 4:} \it Onset of contraction} ($e_R ^2 (R_L) > 
e_L ^2 (R_L) - e^2 > 0$).  When the planet's $e$
decreases below $e_L (R_L)$, $\dot R_p$ and $\dot e$ attain the same
sign such that the planet contracts  despite the tidal dissipation (see
eq. \ref{eq:ttaurr}).  Although $R_L > R_c$ so that the Roche-lobe
overflowing planets are unstable, $R_e < R_L$ for $e < e_L (R_L)$,
{\it i.e.} the tidal dissipation can no longer supply adequate flux of
energy to counterbalance the radiative losses.  In this limit, the
planets undergo contraction.

The initial contraction proceeds on a time scale $\vert \tau_R \vert >
\tau_e$.  The reduction in $e$ and $R_p$ quickly break the thermal
equilibrium such that ${\cal L} > > -\dot E_{\rm tide}$.  In this limit,
the contraction time scale
\begin{equation}
\tau_c (R_p) = \left( {e_R \over e_L} \right)^2 \tau_e
= { q_p G M_p ^2 \over {\cal L} R_p} \tau_e \sim 1.6 \times 10^8 q_p \left( 
{2 R_J \over R_p } \right)^4  \left( {M_p \over M_J} \right)^2 {\rm yr}
\label{eq:taucon}
\end{equation}
for a $M_p =0.63 M_J$ planet (see eq. \ref{eq:taucool}).
Equation (\ref{eq:lessel1}) indicates that the outflow is quenched by the
contraction of the planets' $R_p$ despite a reduction in their $R_L$
as a consequence of $a$ reduction.  Setting $\dot M_{L1} =0$ in
equation (\ref{eq:lessel2}), the conservation of specific angular momentum
$a (1-e^2)$ implies that $\dot a$ has the same sign as
$\dot e$ and that the subsequent circularization leads to orbital
decay.

Since the planet's $e$ is damped before its size is significantly
deflated, the duration of stage 4 is $\tau_4 \sim \tau_e (R_L)$ which
is also comparable to both $\tau_2$ and $\tau_3$ and the age of
typical classical and weak-line T Tauri stars.  For $a < a_{RL}$, $e_R
(a, R_L) < e_L(a, R_L)$, $\tau_R$ remains to be larger than $\tau_e$
as $e$ diminishes well below $e_R$.  In this case, the planet retains 
its extended $R_p (\sim R_L (a_L) \> 10 R_J)$ until its $e$
vanishes  on a time scale $\tau_e (a_t, R_L)$ (see eq. \ref{eq:eovere}).
The asymptotic value of the semi-major axis is $a \sim a_t ( 1 - e_t
^2)$.  The planet's extended size and aspect ratio ($\sim 0.1$)
makes it  detectable with transit searches among young stellar
objects.  For $a> a_{RL}$, $e_L (R_L) > e_R (R_L)$ and the initial
contraction evolves into a rapid deflation to $R_c$ when $e$ decreases
below $(e_L^2 - e_R^2 )^{1/2}$.  The fractional $e$ reduction is
$\delta e/ e_L (R_L) \simeq e_R (R_L) ^2 /2 e_L (R_L)^2 < 1$ with a
similar change in $a$.  In this case, the duration of stage 4 is $\sim
(e_R ^2 (R_L)/ 2 e_L ^2 (R_L) \tau_e$.

{{\bf Stage 5:} \it Runaway deflation and return to stable
equilibrium} ($ e_L ^2 (R_L) - e^2 > e_R ^2 (R_L)$).  The $e$ damping
reduces the energy dissipation rate and the planet's equilibrium radius
$R_e$ to be less than $R_c$ or $\sim 2 R_J$.  For $a < a_{RL}$, $e$
vanishes on a time scale $\tau_e (R_L) < < \tau_R (R_L) \sim \tau_c
(R_L)$.  Thereafter,  the planet contracts  from the Roche radius to $R_c$
on a time scale $\tau_c (R_c)$.  Although planets located at $a >
a_{RL}$ may retain a substantial $e (\sim e_L (R_L))$ when their $e_L
^2 (R_L) - e^2 > e_R ^2 (R_L)$, they also contract to $\sim R_c$ on
the time scale of $\tau_R \sim \tau_c (R_c)$.  But, since $\tau_e
\propto R_p^{-5}$, the $e$ damping time scale for these planets
increases rapidly with their contraction.

Since $e_R (R_p) \propto R_p ^{-1/2}$ (eq. \ref{eq:er}), and $e_L (R_p)
\propto R_e^{(\gamma-5)/2}$ (eq. \ref{eq:eL}), the magnitude of both
$e_R$ and $e_L$ increase by the contraction of $R_p$ when $\gamma <5$
or $R_p > R_c$.  As $e$ becomes smaller than $e_R$ and $e_L$, the
effect of tidal heating diminishes monotonically.  In both limits, the
runaway deflation reduces $R_p$ below $\sim R_c$ on a time
scale $\tau_c (R_c) \sim 10^8 $ yr which is the duration of stage 5,
$\tau_5$.

After  $R_p$ is reduced to $\sim 2 R_J$, the planet's contraction
slows down.  Initially, ${\cal L} > > - \dot E_{\rm tide}$ 
so that the primary source of energy for
the planet's luminosity is the release of gravitational energy.
But, this energy release rate decreases
with time (Burrows et al. 1997).  If $e$ is damped to a negligible
magnitude, $R_p$ would contract to become comparable to that of
Jupiter today.  The retention of any residual $e$ would provide an
additional energy source such that the contraction would continue
until an energy equilibrium is established and $R_e$ attains the
values given by setting ${\cal L} = -\dot E_{\rm tide}$ in
equation (\ref{eq:equalL}).

Associated with this energy dissipation, the planets' $e$ would be
dissipated on a time scale $\tau_e (R_e)$ (see eq. \ref{tau_e}).  But
for relatively large $a$'s and small $R_p$, $\tau_e$ is likely to be
much longer than the expected life span of typical solar type stars
and $e$ is not significantly damped after stage 4.  Similarly, the 
evolution of the planets' $a$ is also likely to terminate. 

\subsection{Semi-major axis expansion and mass reduction factor}
The final dynamical properties of  a tidally inflated planet are
determined by the conservation of total angular momentum.  The
above discussions indicate that unstable planets with $a_i < a_{L2}$
expand to overflow their Roche lobe via both the L1 and L2 points.
The planet's angular momentum is absorbed by the lost mass exterior to
the orbit and it  become tidally disrupted {\it in situ}.  But,
unstable planets with $a_i > a_{L2}$ overflow their Roche radius via
the L1 point and they must absorb all the initial angular momentum.
In this case, the planets' mass loss  is terminated during stage 3 when they
attain $a_t$ and $e_t$.  In the limit that $e_t < < e_i$, the
fractional loss of mass $\Delta M_p / M_p \simeq (a_{t} - R_{i})/2
a_{L} + e_i^2 /(1-e_i^2) \sim 0.05-0.3$ (see eq.  \ref{eq:mpatet}).
Most short-period planets are fractionally less massive than the more
distant planets, though this trend is probably due to observational
selection effects.  Nevertheless, the absence of very massive planets
among them may be indicative of some mass loss for the short-period
planets.

In the above discussions, we show that all planets with $P > P_{eL}
(e=1)$ are stable against runaway tidal inflation.  Also no planet can
continue to overflow its Roche lobe with $P>P_L(e=1)$.  Nevertheless,
$P_L (a_t, e_t)$ can be considerably larger than the circularization
period for isolated planets with similar kinematic properties.
Equation (\ref{tau_e}) indicates that the decay time of $e$ would be
extremely short owing to large radii ($=$ Roche radii) during overflow
phases: in the case that the Roche radius $R_L=4.92R_J$ at $a=0.04$ AU
at the beginning of overflow, equation (\ref{tau_e}) gives $\tau_e$ as
short as $\sim 10^5$ years. When the planets' mass loss through Roche
lobe overflow is terminated at $a_L$, $\tau_e (a_L, R_L)$ is still
less than 1 Myr (see eq.  \ref{eq:tealrl}).  Unstable planets with $a >
a_{RL}$ may retain their residual eccentricity.  Thus, around
stars of similar ages, we expect the coexistence of planets with
periods up to 2-3 weeks, some with initially highly eccentric orbits
but which migrated outward and therefore currently have
circular orbits, and others with orbits in which the planets did
not inflate too much and have stayed near their original
orbits, with initially and currently modest eccentricity.

In the absence of any tidal heating, protoplanets with $R_p \simeq 2
R_J$ overflow their Roche radius when their $a$ is reduced to $\sim
0.014$ AU.  \cite{tri98} suggested that after their tidal interaction
with their protoplanetary disks brings them to such a distance from
their host stars, protoplanets lose mass primarily through the L1
point.  They also proposed that the angular momentum transferred from
the lost mass to the remaining protoplanet may be adequate to
compensate for the angular momentum loss to the disk.  Since this
halting mechanism can only occur when the protoplanets overflow their
Roche lobe, it is unlikely to be effective for planets with periods
greater than 3 days, even taking into consideration their inflation by
tidal dissipation.  For protoplanets which migrated to such close
range, this mechanism alone cannot sustain a protracted angular
momentum drainage by a long-lived disk.  In the absence of a strong
tidal interaction with rapidly rotating host stars or the existence of
a cavity in the inner regions of their disks to halt their migration
(Lin et al. 1996), all protoplanets that migrate to the close
neighborhood of their host stars would perish through tidal
disruption.

In view of the uncertainties in the magnitude of $Q^\prime _p$, we
briefly discuss the dependence of various quantities on it.  The
quantities which linearly vary with $Q^\prime _p$ are the time scale
of eccentricity damping and tidal inflation.  A relatively large
$Q^\prime _p$ can significantly prolong the process of tidal
inflation.  But, the dependence of the critical eccentricity ($e_L$)
for runaway inflation on $Q^\prime _p$ is more modest.  For
sufficiently large $e$'s, the critical values of $a$'s for the onset
of tidal inflation, Roche-lobe overflow, and deflation are relatively
insensitive to the magnitude of $Q^\prime _p$.  Since the mass
reduction factor is determined by the stalling semimajor axis,
it too is an insensitive function of $Q^\prime _p$.  Thus, provided
the magnitude of $Q_p ^\prime < 10^8$, our overall conclusion that
runaway inflation may lead to Roche lobe overflow and the disruption
or the displacement of ultra short-period eccentric extrasolar planets
does not sensitively depend on the equilibrium tidal prescription we
have adopted.

\section{Numerical Simulation of Tidal Inflation}

The analytic approximations  presented in the previous sections  are  useful
for categorizing various potential evolutionary paths.  However, the
stability criteria in equations (\ref{eq:er}) and (\ref{eq:elstable})
neglect the planets' $e$ damping during their evolution towards a
state of thermal equilibrium.  The discussion in the previous sections 
also indicates  that depending on the planets' initial conditions, there
are many possible evolutionary paths and that therefore some of the 
analytic approximations may not be appropriate.

In this section, we present numerical solutions to illustrate the
tidal evolution of ultra-short period planets.  For this quantitative
analysis, we consider a simple extrasolar system which consists of a
solar-mass star and a Jovian planet with an initially  eccentric orbit.
The equations governing the time evolution of $e$, $R_p$, $\Omega_p$
and $a$ involve total energy and angular momentum conservation of the
whole system consisting of a giant planet and a sun-like host star.
Some simplification has been made in our approach.  We do not perform
the calculation for the expansion/contraction of a planet that is
placed in the Roche lobe potential. Instead, we use the evolution code
of BLM for a spherically symmetric, gravitationally isolated giant
planet to estimate the evolution of the planet's  radius.  We include the
heating from both tidal dissipation (appropriate for planets with $a$
ranging from 0.02 AU to 0.05 AU) and the stellar irradiation.  Several
evolutionary calculations are carried out for each value of $M_p$, for
different (constant with time) values of dissipation rates. In each
case the evolution is calculated up to a time of several Gyr or until
the radius becomes constant at the value corresponding to thermal
equilibrium, that is, when the luminosity generated by dissipation
equals the luminosity radiated from the interior.  From these
calculations we generate a table that supplies the relation $\dot
R_p=f(\dot E_{\rm tide},R_p,M_p)$.  Then we can interpolate/
extrapolate the values of $\dot R_p$ from the table by entering it
with any dissipation rate $\dot E_{\rm tide}$ calculated from either
the equation of $e$-damping (eq. \ref{eq:etide} with the
synchronization condition: $n=\Omega_p$), or that of synchronization
damping (eq. \ref{eq:etide} with the condition for a circular orbit:
$e=0$).  We do not run the simulation with both synchronization and
circularization heating at the same time. The heating due to
synchronization is much smaller than that due to circularization, as
roughly estimated in equation (\ref{eq:tauro}).  This point will be justified
by our simulation results in \S6.2. In summary, we solve equations
(\ref{eq:edot0_e}), (\ref{eq:etide}), (\ref{eq:lessel2}), (\ref{eq:dotr2c}),
and (\ref{eq:dotOmega}) together with the interpolation/extrapolation function
$\dot R_p=f(\dot E_{\rm tide},R_p,M_p)$ for time evolution of
$e$, $\Omega_p$, $a$, $R_p$, $M_p$, and $\dot E_{tide}$.

Several cautions need to be addressed with our approach. First
of all, since the simulation carried out by BLM does not take
into account the planet spin, we do not include the energy $\dot
E_{spin}$ to calculate $\dot R_p$.  
In \S3.1, we already showed that $E_{spin}$ is smaller than
$E_{bind}$, and therefore ignoring $E_{spin}$ should not
influence the internal structure too much.
The second caution is that the value of $\dot R_p$ in the tables are
computed using constant heating rates. In principle, the evolution of
stratification of planets due to thermal imbalance in the case of
quasi-hydrostatic equilibrium should react differently to different
heating methods. In \S3.2, we suggest this assumption might be
appropriate
because much of the gas in giant planet interiors has a highly non-ideal  equation
of state so that the tidal dissipation mainly leads to temperature
increases without any significant changes in the pressure.

Mass loss starts to occur once the photosphere of an inflated planet
reaches its Roche radius.  Overflow gas would redistribute mass and
angular momentum of the whole star-planet binary system and then alter
the semi-major axis with time. In our numerical calculations, we
assume the planets overflow their Roche lobe via the L1 point only and
that the system evolves in a conservative manner. We employ
equation (\ref{eq:dotr2c}) to estimate the mass loss rate $\dot M_{L1}$.
Our simulations are terminated once $\dot M_{L1} >0$. 
This situation usually occurs when the planet
starts to contract gravitationally due to inefficient tidal
dissipation as $e$ drops to low values in our simulations.

\subsection{Eccentricity Damping}
Table~\ref{tab:e_roche} shows the minimal $e$'s required for a giant
planet to reach its Roche lobe.  This $e$ is equivalent to $e_L$ in
equation (\ref{eq:eL}).  While the larger initial planet radii ($R_{p,
init}=1.8R_J$ for $1M_J$ and $R_{p, init}=2.0R_J$ for $0.63M_J$)
represent earlier phases of giant planets, the smaller initial radius
($R_{p, init}=1.2R_J$) is used as a later phase of a giant planet that
already contracted and is reaching thermal equilibrium.  In all these
models, we set $M_\ast= 1 M_\odot$ and $Q_p ^\prime = 10^6$.  
The data in parentheses are the times  taken
for the planet to expand from the initial conditions to fill its Roche
lobe in units of Myr. This time scale is equivalent to $\tau_R$ in
equation (\ref{eq:taur1}).  The table shows that this time scale increases
with $a$, resulting partly from larger $\tau_e$ as indicated in
equation (\ref{tau_e}) (i.e.  $e$ decays more slowly at larger $a$ due to a
weaker tidal effect)  and partly from larger $R_L$ at larger $a$
(i.e.  a planet requires a  longer time to reach the larger Roche lobe at larger $a$).
The $e$'s for small-size planets are higher than those for large-size
planets as a result of larger gravitational binding energy for the
planets starting from smaller radii.  The same reason explains why
planets with a core require larger $e$'s to reach their Roche radius
than planets without a core. The results of our simulations  indicate
that a planet of $0.63M_J$ without a core (with a core) at 0.04 AU
starting at the critical eccentricity 0.140 (0.156) and 2 $R_J$ will
reach its Roche radius in about 1 Myr. This time scale is shorter
than the one given by equation (\ref{eq:taur1}), which can be explained by
the steep change of $e$ and $R_p$ as $R_p$ approaches $R_L$ and the
reduced gravitational binding energy that needs to be overcome 
for larger radii.  

The results in Table~\ref{tab:e_roche} also show the value of the
minimum $e$ is inversely proportional to $R_p$ which is in agreement
with the expression in equation (\ref{eq:eL}).  The weak dependence with
$M_p$ is probably associated with ${\cal L}_o$ being an increasing
function of $M_p$.  Although planets located at larger $a$'s require
larger values of minimal $e$'s and thereby survive more easily, the
minimum value of $e$ increases  less rapidly with $a$ than that
expressed in equation (\ref{eq:eL}).  The minimum values of $e$'s are
generally larger than $e_L$, especially for small $a$.  For $a<
a_{RL}$, the more appropriate condition for planetary inflation is $e
> e_R$ (see eq. \ref{eq:aRL}).  The numerical results are in good
agreement with the expression of $e_R$ in equation (\ref{eq:er}).

Figure~\ref{fig:uptoRoche} illustrates the time evolutions of $e$,
$R_p$, $a$, and $\dot E_{\rm te}$ for a coreless planet with an original
$M_p=0.63M_J$, initially located at 0.04 AU (solid lines) and
0.06 AU (dashed lines).  Starting with $e\approx 0.14$ ($e\approx
0.26$), the planet at $a=0.04$ AU ($a=0.06$ AU) spends $t\approx 1.4$
Myrs ($t\approx 8.2$ Myrs) and then is able to expand up to its $R_L$
at which ${\cal L}_e= -\dot E_{\rm te}$.  After this point, the planet
will contract since the dissipation rate $\dot E_{\rm te}$ has already
decreased with time before the planet's size reaches its $R_L$ as is
shown in the lower-right panel of Figure ~\ref{fig:uptoRoche}. In the
case for larger values of $a$ such as 0.06 AU, $\dot E_{\rm te}$
starts to increase more slowly than that for smaller $a$ as a result
of a longer damping time scale of eccentricities, leading to a slower
rate of expansion as shown in the upper-right panel in
Figure~\ref{fig:uptoRoche}. As long as the initial $e_i$ is large enough
for the planet to expand slowly through the initial phase and then is
able to get larger, the inflation rate is subsequently enhanced by its
larger radius ($\dot E_{\rm te} \propto R_p^5$) and smaller gravitational
binding energy.  The lower-left panel in Figure~\ref{fig:uptoRoche}
shows that $a$ decreases with time due to circularization. The
critical eccentricity 0.14, calculated by the detailed simulation for
a planet of $M_p=0.63M_J$ at $a=0.04$ AU as shown in
Table~\ref{tab:e_roche} and in Figure~\ref{fig:uptoRoche}, just
coincides with the value $e_R=0.14$ based on the rough estimate as
shown in equation (\ref{eq:er}), and is larger than the value $e_L=0.06$
that is estimated based on equation (\ref{eq:eL}) with an extrapolation of
BLM's data as described in \S3.2.

Based on the physical process that planet mass is lost through L1 (see
eq. \ref{eq:dotr2c}), we calculate the smallest initial $e_i$'s required
for a giant planet to move from its initial orbit with a semi-major
axis $a_{init}$ to its final orbit with a semi-major axis
$a_{final}$. The results without a core and with a core are shown in
Tables~\ref{tab:e_crit_1.8}, \ref{tab:e_crit_1.2}, and
\ref{tab:e_crit_63}. The mass loss rate is calculated based on
equation(\ref{eq:dotr2c}).  The data for $\lambda =1$ and $\lambda =3$ are
very close to the data for $\lambda=2$. Therefore we only show the
results for $\lambda=2$ in this paper.  Giant planets move away from
their host stars by receiving orbital angular momentum transfered to
them from the overflow gas, and their masses are reduced by a fraction
before they become detached from their Roche lobes. The data in bold
face as displayed in parentheses show the final masses of giant
planets in units of $1M_J$ after their radii become detached from
their Roche radius.  Typically, $M_p$ is reduced by $\approx$ 15\% to
40\% for planets with initial $e$'s comparable to the critical
values. The surviving mass is expected to be smaller if the planets'
initial $e$'s are larger than the critical values.  These minimal
$e$'s depend on the initial conditions: the greater the initial
gravitational binding energy (due to smaller radii, larger mass, or
the existence of a core), the greater the dissipation, and therefore
the larger the $e$, is  needed to inflate the planet to attain the
same final condition.  For an outward migration from $0.03$ AU to
$0.04$ AU, a planet starting with $1M_J$ and $1.2R_J$ requires
$e\gtrsim 0.2$, a planet initially with $1M_J$ and $1.8R_J$ needs
$e\gtrsim 0.17$, and a planet with $0.63M_J$ and $2R_J$ should start
with $e\gtrsim 0.14$. However, a giant planet of $1M_J$ and $1.8R_J$
originally located at 0.04 AU requires $e\geq 0.204$ (without a core)
or $e\geq 0.223$ (with a core) to migrate to 0.05 AU, indicating the
trend of difficulty for a giant planet located further away to move
outwards by Roche lobe overflow.  The difference of critical $e$'s
between a planet with a core and without a core increases with $a$
because weaker tidal interaction occurs at larger $a$.  The first
values in parentheses (not in bold face) show the time scale in units
of Myrs for $a$ from $a_{init}$ to $a_{final}$. The time scales range
from some small fraction of one Myr to more than 10 Myrs depending on
initial values of $e$, $a_{init}$, and $a_{final}$.

Figure~\ref{overflow} displays the time evolution of $e$, $R_p$, $a$, 
$M_p$, and $\dot E_{te}$ for a planet without a solid core. The
initial radius is $1.8$ $R_J$ and the initial mass is 1
$M_J$. Starting with the $e=0.155$ which is the minimal eccentricity
to get to the final orbital distance 0.04 AU as shown in
Table~\ref{tab:e_crit_1.8}, the planet reaches its Roche lobe around
$t=0.34$ Myrs, marked by dashed lines in the figure. The beginning of
the Roche overflow phase is illustrated in the $M_p/M_J$ vs $t$ plot
where $M_p$ starts to fall off around $t=0.34$ Myrs. At the same time,
$\dot E_{te}$ reaches the maximal value.  Subsequently, the planet
loses its mass, moves outwards, and increases its size as a result of
Roche lobe overflow. The decrease in $\dot E_{te}$ during the overflow
phase is a consequence of the increase in $a$.  The mass loss rate is
determined by equation (\ref{eq:dotr2c}) with the condition that $\lambda
=2$.  After having lost 0.16 $M_J$, the planet starts to contract and
therefore detaches from its Roche lobe at $a\approx 0.04$ AU due to
inefficient $\dot E_{\rm te}$ (and $e$) when $t\approx 0.45$
Myrs. Note that $a$ slightly decreases with time before the beginning
of the Roche phase ($t\lesssim$ 0.34 Myrs) as a result of orbital
circularization. However, the change of $a$ is dominated by the effect
of overflow during the mass-losing phase (i.e. $|\dot a_m|>|\dot a_e|$
in equation (\ref{eq:dota}) when $t>$ 0.34 Myrs).

Figure~\ref{overflow2} illustrates the evolution of the same planet
as shown in Figure~\ref{overflow} but with a larger value of initial
eccentricity. With a large value of initial $e$, the whole dynamical
evolution should more or less go through a sequence of the stages
as illustrated in \S5.
Owing to the larger initial $e=0.2$ shown in 
Figure~\ref{overflow2}, 
the planet loses more mass and therefore migrates even farther out 
than the case (as shown in Figure~\ref{overflow}) with the initial $e=0.155$.
The planet enters its Roche-overflowing phase at $t\approx 0.13$ Myrs.
During the period $t=0.13$ to 0.2 Myrs ({\it i.e.} the early phase of
the Roche overflow), $R_p$ increases roughly from $4R_J$ to more than
$7R_J$ but $e$ declines approximately from $0.14$ to $0.12$, demonstrating
``Stage 2'' described in \S5. The following phase after $t>0.2$ Myrs
featured by the flattening trend of $R_p$ with time confirms the
stagnation stage ({\it i.e.} Stage ``3'') described in \S5.
The final contraction stage is not able to be simulated with
our method which is based on extrapolation/interpolation of
the planet data inflated up to the equilibrium radii.

The above results are based on the assumption that overflow occurs
only through the L1 point. However, there might be an additional
concern in the star-planet interacting system. The gravitational
potential difference between the L1 and L2 points is drastically
reduced for the small mass ratio $M_p/M_\ast \approx 10^{-3}$,
therefore raising the possibility of overflow through the outer
Lagrangian point (L2). Our simulation for a planet of $1M_J$ without a
core at $a=0.03$ AU shows that for the initial eccentricity $e=0.155$
($e=0.5$), the rate of mass transfer across equipotential surfaces is
about $\dot M_a \approx 4\times 10^{17}$ g/s ($9\times 10^{18}$ g/s),
which should be roughly comparable to the mass flux via the L1 point.
This flux is given by equation (\ref{eq:mdl1}).  Let us estimate this mass
flux $\dot M_{L1}$ based on the conditions at the photosphere:
$\rho=10^{-8}$ g/cm$^3$, $c_s=4\times 10^5$ cm/s.  These values would
give rise to the mass loss rate $\dot M_{L1} \approx 10^{18}$
g/s. Hence $\dot M_{L1} \gtrsim \dot M_a$ for initial $e=0.155$, but
$\dot M_a \gtrsim \dot M_{L1}$ for initial $e=0.5$. As we discuss in
\S4, mass should be primarily transferred via the L1 point in the
small-$e$ limit since the bottom of the planet's photosphere
underfills the L1 point and the density scale height $h_{\rho}$ is smaller
than $\Delta D$ as shown in equation (\ref{eq:hoverd}).  In the large-$e$
regime, the bottom of the planet's photosphere overfills L1, probably
leading to a significant amount mass loss through L2. If the planet
starts at $a=0.02$ AU, our simulation shows that $\dot M_a
\approx 5\times 10^{18}$ g/cm$^3$ for initial $e=0.25$, therefore
increasing the chance of overfilling the L1 point as indicated by
equation(\ref{eq:rhol1a}) for smaller $a$.

In the previous section, we suggested that in  a system with $2ea <
h_\rho$, mass transfer is continuous because the gas removed from the
atmosphere is replenished on a dynamical time scale by the requirement
of hydrostatic equilibrium.  For planets with larger $e$'s, the
envelope as well as the atmosphere needs to adjust.  Also  as reasoned in
the previous section, the underlying envelope of the planet expands at
a rather slow rate, which should not be of importance in determining
$\dot M$ if there is continuous mass overflow in one orbital
period. Our simulation shows that the fastest expansion rate of the
planets' envelope is $\dot R_p \sim 0.1$ cm/s for an initial $e=0.5$,
and $\sim 0.01$ cm/s for an initial $e=0.155$ in the case of a giant
planet of $1M_J$ without a core at $a=0.03$ AU.

When a planet moves from perihelion to aphelion, its Roche
radius increases at a rate
\begin{equation}
\dot R_L \approx \left( {M_p \over 3 M_\ast} \right)^{1/3} 
\left( {2ae \over P/2} \right)
\sim 10^5 \left( {M_p\over M_J} \right)^{1/3}
\left( {M_\ast \over \msun} \right)^{1/6}
\left( {0.04 {\rm AU} \over a } \right)^{1/2}
\left( {e\over 0.2} \right) {\rm cm/s},
\end{equation}
which is indeed much larger than the expansion rate $\sim 0.1$ cm/s due to
the $e$-damping from our simulation. On the other hand,
\begin{equation}
2ae \approx 1.66\times 10^9 \left( {e\over 0.2} \right)
\left( {a\over 0.04\,{\rm AU}} \right) \,{\rm cm}
\sim h_\rho \approx 2-3\times 10^9 \,{\rm cm},
\end{equation}
where $h_\rho$ is evaluated at $0.04$ AU (see \S4.2).  As a result,
one would expect mass overflow via the L1 point to be sustained in
different phases of one orbital period despite the re-expansion of
planet's atmosphere, although the most prominent overflow occurs
around the perihelion. Hence we conclude that using the Roche radius
based on perihelion is a reasonable approximation for small $e$'s.
However, $h_\rho$ would be much larger than $2ae$ at smaller $a$ since
$h_\rho$ gets larger due to more intensive stellar radiation and
furthermore $2ae$ gets smaller.


So far we have not considered the effect of the planet's tidal
interaction with the protostellar disk and the consequent inward migration. 
The primary concern about inward migration is that 
if the planets' orbits
are almost circularized by the time they reach 
$a=0.04$ AU, tidal dissipation within them wouldn't be available to
inflate their radii. As has been discussed in \S3.3, there might be
a couple of theoretical mechanisms to excite/damp $e$ as giant planets
migrate in through a disk. We do not include these uncertain mechanisms 
in this paper, but only take into account the damping of
$e$ due to $\dot E_{te}$ as a function of various inward migration rates, using 
 a simple prescription $a/\dot a=$constant.  Figure
~\ref{edecay} shows the evolution of $e$ of a coreless synchronized
planet, with $1M_J$ and $1.8R_J$, moving toward its host star from
$a=0.4$ AU on three different time scales: $-a/\dot a=$5,
1, and 0.1 Myr.  The
simulations start with $e=0.2$ and $e=0.3$ and end before the planet
reaches its Roche lobe. As shown in Figure~\ref{edecay}, $e$'s do not
drop noticeably from their initial values until the planet reaches
$a=0.04$ AU; thus at this point enough energy from orbital
circularization still remains to inflate the planet to its Roche
radius.  The orbital distance $a$ for Roche overflow shifts to smaller
values when $a/\dot a$ or initial $e$ is smaller.

The minimal $e$'s listed in the first column (i.e. those data with the
initial condition $a_{init}=0.02$ AU) in Table~\ref{tab:e_crit_63} are
unexpectedly large compared with those in the first columns in
Tables~\ref{tab:e_crit_1.8} and \ref{tab:e_crit_1.2} when the values
of these minimal $e$'s are $\gtrsim 0.2$.  In principle, an
internally-heated planet starting  with a smaller mass and a larger size
should overcome its gravitational binding energy more easily. In
reality, this surprising outcome results from a special initial
condition for any large $e$: the initial radius $R_p=2R_J$ is
already larger than the planet's Roche radius $R_L$ for $R_p=0.63M_J$,
$a=0.02$ AU, and $e>0.2$ according to equation (\ref{eq:roche}).  Under such
a circumstance, the planet cannot be inflated to a large size before
overflow happens, leading to less intensive tidal heating and
therefore requiring a larger $e$ to reach $a=a_{final}$. In
contrast, a planet at $a_{init}=0.02$ AU starting with more mass such
as $1M_J$ can be inflated to a larger size before overflow starts to
take place, resulting in more intensive tidal heating and thereby
requiring a smaller $e$ to migrate to the same final location
$a_{final}$.

\subsection{Synchronization}
In this subsection, we shall show that unlike orbital circularization,
synchronization is not a promising source of dissipation for a planet
within 0.04 AU of its host star since the planet is almost
completely synchronized during the course of its inward migration. In
this analysis, we consider the dissipation solely due to
synchronization, but including orbital migration and the planet's
inflation/contraction. There will be three equations governing the
time evolution of $\Omega_p$, $R_p$, and $a$. We employ the same
assumptions and approach for evaluating $\dot R_p$ and $\dot a$ for
synchronization damping as we described in the previous section for
eccentricity damping.  In the case of synchronization for a circular
orbit, the dissipation is given by $\dot E_{{\rm t} \Omega}$ in
equation (\ref{eq:etide2}):
\begin{eqnarray}
\dot E_{{\rm t} \Omega} =
{I_p |n-\Omega_p | \Omega_p \over \tau_\Omega }
=2.41 \times 10^{-5}
\left( {R_p \over R_J} \right)^5
\left( {0.04\,{\rm AU} \over a} \right)^{9/2}
\left( {M_\ast \over \msun} \right)^{3/2}
\left( {10^6 \over Q_p} \right) \nonumber \\
\left[ 0.34 \left( {M_\ast \over \msun} \right)^{1/2}
\left( {0.04\,{\rm AU} \over a} \right)^{3/2} -
{1\, {\rm day} \over 2\pi/\Omega_p} \right]^2 \Lsun. 
\label{L_sync}
\end{eqnarray}

In equation (\ref{eq:omedot}), $\dot \Omega_p$ contains two pieces, each of
which is associated with a different physical process: synchronization
based on the conservation of orbital and spin angular momentum
(e.g. see \cite{md99}), and inflation/contraction based on the
conservation of spin angular momentum such that
\begin{equation}
\dot \Omega_p=\dot \Omega_{tide} + \dot \Omega |_{I_p\Omega={\rm constant}}
=\left( {\rm sign}(n-\Omega_p)
{1 \over \tau_{\Omega}} - {\dot I_p \over I_p} \right)\Omega_p.
\label{eq:dotOmega}
\end{equation}

Figure~\ref{sync} shows the time evolution of $a$, $\Omega_p$, $r_p$,
$\dot E_{\tau \Omega}$, and the synchronization factor $\Omega_p/n$,
for a 0.63 M$_J$ planet with three different migration
rates $a/\dot a=$ 5, 1, and 0.1 Myr. Initially the planets are fast
rotators at a spin period $2\pi/\Omega_p =2$ days, and they start from
0.4 AU with $R_p=2R_J$. This corresponds to the condition
$\Omega_p/n=46.48$. Then they migrate until they overflow their Roche
radii.

In the beginning the spin periods $2\pi/\Omega_p$ decrease as the
planets move in as a result of contraction. The contraction phase
continues until the tidal interaction becomes so important that the
spin frequencies $\Omega_p$ are slowed down by the orbital frequencies
$n$ which increase with time due to migration.  After $n$ catches up
with $\Omega_p$ (this occurs around 0.05--0.1 AU depending on
migration rates) and then becomes larger than $\Omega_p$ as a result
of continuous migration, the spin periods decrease again mainly due to
the fact that $\Omega_p$ tries to catch up with $n$ because of
synchronization.  As a result there are two dissipation peaks due to
synchronization during the course of migration as illustrated on the
$\dot E_{t \Omega} /L_{\odot}$ diagram in Figure~\ref{sync}.  The
first dissipation peak happens at a relatively large $a$, when $n$
catches up with $\Omega_p$, and the second peak occurs at a smaller
$a$ when $\Omega_p$ catches up with $n$. The second peak has a
relatively small amplitude compared to the first peak because the
difference between $n$ and $\Omega_p$ is small during the second
dissipation peak.  Therefore, one can essentially state that the
system has reached its synchronous state after $n$ catches up with
$\Omega_p$ at around 0.05 AU.  This result is in agreement with the
discussion in \S3.2.

The continuous contraction as shown in the upper-right panel in
Figure~\ref{sync} is just a result of small $\tau_{t \Omega}$ shown in
the lower-left panel in Figure~\ref{sync}.  A fast migration rate
$a/\dot a=$0.1 Myr is unable to boost tidal dissipation high enough to
inflate the planet.  The final Roche lobe overflow taking place at
$a\approx$0.015 AU (0.016 AU for $-a/\dot a=$ 0.1 Myr, 0.015 AU for
$-a/\dot a=$1 Myr, and 0.014 AU for $-a/\dot a=$ 5 Myr) is therefore
not due to radius expansion, but due to a decrease of the Roche radius 
as the planet migrates  in.  Although the simulation is carried out
only for a coreless planet with a relatively small mass $M_p=0.63M_J$,
planets with cores or of larger masses such as $M_p=1M_J$ are more
difficult to inflate to their Roche radii by synchronous
dissipation due to their larger gravitational binding energy.

We also plot the evolution of a coreless planet with $0.63M_J$ which
is being tidally heated as a result of synchronization, starting
without any orbital migration. The result is illustrated in
Figure~\ref{sync_0.03AU}. The planet starts as a rapid rotator with
$\Omega_p/n \approx 7.64$ at $a=0.03$ AU (i.e.  $2\pi /\Omega_p =
0.25$ days). This value of $\Omega_p$ has been chosen to be
sufficiently large for the generation of a substantial amount of $\dot
E_{t\Omega}$, but small enough to limit $E_{\rm spin}$ to be less than
$E_{\rm bind}$.  The synchronization time scale is extremely short:
$\tau_\Omega \sim 1000$ yrs according to equation (\ref{eq:tauo}) for
$M_p=0.63M_J$, $a=0.03$ AU, $R_p=2R_J$, and $2\pi/\Omega_p=0.25$ days.
Figure~\ref{sync_0.03AU} shows that $\Omega_p /n \rightarrow 1$ 
within a few thousand years.  Consequently, even though the
equilibrium radius for the intense initial dissipation rate ($0.01
L_\odot$) is $R_p = 2.9 R_J$, the planet exhausts its energy source
and stops its expansion in a couple of thousand years, well before
reaching its Roche lobe.  This result confirms the general statement
$\tau_{R\Omega} > \tau_\Omega$ as described by equation (\ref{eq:tauro}).

\section{Summary and Discussion}

Based upon interpolation/extrapolation of numerical results of
internal structure by BLM, we have demonstrated, analytically and
numerically, that giant planets with masses $\le 1M_J$, initial radius
$\approx 1.8R_J$, modest $e$'s ($\gtrsim 0.1-0.22$), and small $a$'s
($< 0.04$ AU) are likely to be tidally inflated beyond their Roche
lobe as they undergo $e$ damping. These conditions may be slightly
modified by their internal structure (with or without cores) and by
inward migration due to disk-planet interaction. The subsequent mass
loss depends on the modification of the internal structure after these
tidally inflated giant planets have lost some of their initial mass.
Therefore this adjustment is difficult to estimate through our
approach based on the numerical data without mass loss by BLM.  The
degree of outward migration will determine whether or not this Roche
lobe overflow model is likely to be responsible for the lack of giant
planets within 0.04 AU.  How much mass the planet will lose and how
far out the planet will migrate is a matter of competition between
several factors. Mass loss increases the expansion rate for a given
dissipation rate. Our simulation shows that the planet radius, in this
case confined by the Roche radius, increases as the planet gains
specific angular momentum from the overflow gas through the L1
point. Increasing the semi-major axis $a$ due to Roche overflow via
the L1 point inevitably reduces the internal tidal heating.  All of
these factors undoubtedly depend on the evolution of the internal
structure under the condition of mass loss. Nevertheless, we adopt a
simplified method by employing the parameter $\lambda$ to relate the
mass loss rate to the expansion rate of the planet (see
eq.(\ref{eq:dotr2c})).  Based on reasonable values of $\lambda$
between 1 and 3, our results indicate that a planet located at larger
$a_{init}$ requires a larger value of critical $e$ to move out to the
same $a_{final}$.  The minimal values of $e$ to reach the Roche radius
increase rapidly as $a_{init}$ increases from 0.03 AU to 0.05 AU as a
natural result of the steep dependence on $a$ for tidal interaction.
This result depends very weakly on the magnitude of the highly
uncertain value of $Q_p ^\prime$.  Therefore, our model suggests that
the lack of planets inside 0.04 AU might be explained by outward
migration due to Roche lobe overflow rather than extreme mass loss.
On the other hand, if the protoplanetary disk is still present, the
outward migration could be countered by the inward migration from
disk--planet interaction; in this case extreme mass loss could
actually provide the explanation. The eccentric orbits $e>0.2$ of
extrasolar Jovian planets are theoretically likely as described in
\S3.3 and are commonly observed,  except near the host star. Also, we
have shown that eccentricity damping before a planet migrates to
within 0.04 AU is so small that the mechanism of eccentricity damping
can provide sufficient energy to inflate a planet to the Roche radius
inside 0.04 AU of its host star. Thus Roche lobe overflow due to
eccentricity damping seems to be a  reasonable mechanism to
explain the apparent lack of giant planets within 0.04 AU based on our
approach.  Around stars of similar ages, we expect the coexistence of
planets with periods up to 2-3 weeks, some with initially highly
eccentric orbits but which migrated outward and therefore
currently have circular orbits, and others with orbits in which
the planets did not inflate too much and have stayed near their
original orbits, with initially and currently modest eccentricity.

The amount of mass loss provides an estimate of the metallicity
enhancement of the host star, provided that (1) the planet's envelope
is substantially enhanced in metals relative to the stellar metal
abundance, and (2) some fraction of the overflow mass is able to
accrete onto the star.  The internal structure of Jupiter and Saturn
is somewhat uncertain (Wuchterl {\it et al.} 2000) ranging from that
with a solid core which contains most of the heavy elements  to those
where metal-enriched gas is well mixed in the convective envelope.
But in both sets  of models, the heavy element abundance inside Jupiter
and Saturn is thought to be 5-10 times larger than that of the Sun.
During the stage 2, if the tidally inflated planets lose most of their
mass via the L1 point, all the tidal debris would be accreted by the
host stars.  But if mass flow occurs through both  the L1 and L2 points,
the tidally inflated planet would be disrupted rapidly with the tidal
debris forming an accretion disk which eventually channels the metal
rich gas onto the host star.  As long as the stars do not have a deep
convective envelope, as is  true in the case of early-type stars or
during the later evolutionary stages of solar-type stars, the effect
of metallicity enhancement should be observable on the stellar surface
(Sandquist et al. 2002).  Further investigation should be carried out
to clarify these effects.

Once giant planets inwardly migrate to $\approx$ 0.04 AU through the
planet-disk interaction, their spin frequencies are more or less
synchronized with orbital frequencies. In spite of the fact that some
dissipation is generated as spin frequencies catch up with orbital
frequencies within 0.04 AU as a result of inward migration, our
results with different inward migration rates show that the internal
tidal damping due to synchronization alone is weak enough that the
giant planets contract during their entire evolution.  A simulation
for a fast-spinning planet at 0.03 AU ($\Omega_p /n \approx 7.64$)
with no inward migration due to disk-planet interaction shows that the
giant planet still cannot be inflated to its Roche radius, consistent
with the relation that $\tau_{R\Omega} > \tau_\Omega$ induced from
equation (\ref{eq:tauro}).  However, planet winds may enhance
synchronization heating if they can carry a sufficient amount of spin
angular momentum out of the planet, thereby holding the planet out of
synchronous rotation.  This effect should be considered in the future,
in a model perhaps analogous to that of the black widow pulsar
(\cite{as94}).

The migration model through the planet-disk interaction alone seems to
have difficulty in accounting for an over-abundance of lower mass
planets at $a<0.1$ AU (\cite{allp}).  Although the over-abundance may  possibly 
be a result of 
observational bias, it could,  particularly
near $a=0.04$ AU, be due to the
retreat of Roche-lobe-filling  planets from the region
inside $a=0.04$ AU. Whether the paucity of the massive close-in
Jovian planets (\cite{zm02})
results from our scenario is left undetermined until more
simulations are carried out for the close-in planets with 
$M_p >1M_J$ (but, see \cite{pr02} and \cite{jiang03}for another model).
The distribution of giant planets at 0.03 AU has been simulated by
\cite{tri98} by taking into account the effect of planet-disk
interaction, the tidal torque from a fast rotating star, and Roche
lobe overflow, but without considering tidal inflation.  The effect of
tidal inflation causes giant planets to reach the Roche lobe phase at
a larger $a$.

The expansion/contraction rates $\dot R_p/R_p$ used in our simulation
are taken from the simulations by BLM which are carried out
for constant heating rates.  But, the
dissipation rates actually increase with time for $e$ larger than
the minimal values as
a consequence of expansion rates larger than eccentric damping rates:
\begin{equation}
{ d \dot E_{te} /dt \over \dot E_{te} }\approx 2{\dot e\over e}+5{\dot R_p
\over R_p}.
\end{equation}
We have used equations (\ref{eq:etide3}) and (\ref{tau_e}) to derive
the above equation. Therefore, the minimal values of
eccentricities required for planets to reach the Roche
limit in our simulations  are likely underestimated. 
However, one thing to keep in mind is that all the
uncertainty of tidal friction is hidden in the dissipation quantity
$Q^\prime _p$, and we have assumed that the tidal heating is uniformly
distributed in planet interiors. Since tidally inflated planets have
radiative envelopes (BLM) and it is the radiative envelopes
that respond most to the deposition of thermal energy, the effect of
tidal expansion might be enhanced due to more dissipation generated in
the radiative envelopes as a consequence of the action of radiative
damping on the dynamical tide (\cite{zahn77}).

    The overflow mechanism that we explore in this paper might offer a
reasonable explanation for the recent HST observation results on a
lack of Jovian planets with orbital periods $\lesssim$ 4 days around
main-sequence stars in the globular cluster 47 Tucanae
(\cite{47Tuc}). In addition to the breakup of planetary systems during
frequent encounters with single stars and binaries in the dense core
of 47 Tuc or the scenario that the lack of giant planets results
from metal-poor circumstellar disks, 
eccentricities of planets can get excited after a close
encounter (\cite{ds01}) with a binary system.  Our simulations show
that the minimal eccentricities required for a giant planet of $1M_J$
starting with $1.2R_J$ to move from $a_{init}=0.04$ AU to
$a_{final}=0.05$ AU is $\approx 0.285$, and that the minimal
eccentricity for the same planet but at $a_{init}=0.05$ AU
to reach its the Roche radius is $\approx 0.3$.
Therefore, with large
eccentricities possibly excited in dense stellar environments, a lack
of the Jovian planets with very short periods in 47 Tuc could result
from the same effect via Roche lobe overflow caused by sufficient
tidal inflation. Further investigation on this topic is ongoing
(\cite{kapsoo}). Tidally-inflated Jovian planets in eccentric
orbits should produce deep transit light curves with shorter
flat bottoms, providing an unique signature for the surveys for
transiting hot Jovian planets in young extrasolar systems
with spectroscopic follow-ups. However, the possibility of
finding these objects in surveys, even conducted in some young 
open clusters, would be small due to the short time scales of
tidal inflation in our model.

\acknowledgements We wish to thank A. Burkert, M. Choi,
I. Dobbs-Dixon, R. Klessen, G. Laughlin, J. Lim, R. Mardling,
G. Novak, K. Oh, E. Vishniac, and H. Yee for useful conversation and
valuable comments. We also thank an anonymous referee for highlighting
the uncertainties associated with the $Q_p ^\prime$ values. Part of
this work was completed when one of us (PG) was a visitor at the
UCO/Lick Observatory, and he is grateful to K.-Y. Lo for the support
of this project. This work is supported by NSF and NASA through grants
AST-9987417 and NCC2-5418.

\clearpage
\begin{deluxetable}{crrrrr}
\footnotesize
\tablecaption{Minimal eccentricities required to reach Roche lobes.
The cases without a core (the values before /)
and with a core (the values after /) are displayed. The 
numbers in parentheses are the time taken from the initial conditions
to the initial Roche phase.
\label{tab:e_roche}}
\tablewidth{0pt}
\tablehead{
\colhead{$M_p$} & \colhead{$R_{p,init}$} & \colhead{0.02 AU} &
\colhead{0.03 AU} & \colhead{0.04 AU} & \colhead{0.05 AU}
}
\startdata
$1M_J$    & $1.8R_J$ & 0.095(0.08)/0.097(0.07) & 0.147(0.49)/0.154(0.41) &
0.203(1.89)/0.220(1.61) & 0.274(4.44)/0.299(3.51) \\
$1M_J$    & $1.2R_J$ & 0.129(0.20)/0.129(0.18) & 0.183(1.64)/0.192(1.07) &
0.252(7.17)/0.284(7.18) & 0.313(12.7)/0.385(12.6) \\
$0.63M_J$ & $2.0R_J$ & 0.055(0.05)/0.055(0.06) & 0.101(0.32)/0.103(0.23) &
0.140(1.35)/0.156(0.93) & 0.195(3.82)/0.228(2.65) \\
\enddata
\end{deluxetable}

\begin{deluxetable}{crrr}
\footnotesize \tablecaption{Minimal eccentricities for planets of
initial $1M_J$ and $1.8 R_J$ to migrate outwards from initial
separations $a_{init}$ to final separations $a_{final}$ based on the
equation (\ref{eq:dotr2}) with the condition that $\lambda =2$.
The first values in parentheses show the time scales in units of
Myrs for $a$ to reach $a_{\rm final}$ from $a_{\rm init}$.
The second values in parentheses, which are in bold face,
show the final masses in
units of $1M_J$ when planet radii detach from their Roche lobe.  The
cases without a core (the values before /) and with a core (the values
after /) are displayed.
\label{tab:e_crit_1.8}}
\tablewidth{0pt}
\tablehead{
\colhead{$a_{\rm final}$} & \colhead{$a_{\rm init}=0.02$ AU} &
\colhead{$a_{\rm init}=0.03$ AU} &
\colhead{$a_{\rm init}=0.04$ AU}
}
\startdata
$\ge$0.04 AU & 0.131(0.19, {\bf 0.70})/0.147(0.15, {\bf 0.70}) &
0.155(0.45, {\bf 0.84})/0.167(0.36, {\bf 0.85}) & \\
$\ge$0.05 AU & 0.145(0.29, {\bf 0.62})/0.170(0.20, {\bf 0.62}) &
0.162(0.45, {\bf 0.76})/0.180(0.34, {\bf 0.76}) &
0.204(1.83, {\bf 0.87})/0.223(1.45, {\bf 0.87}) \\
$\ge$0.06 AU & 0.152(0.59, {\bf 0.57})/0.183(0.47, {\bf 0.57}) &
0.174(0.47, {\bf 0.69})/0.199(0.32, {\bf 0.69}) &
0.207(1.69, {\bf 0.80})/0.229(1.24, {\bf 0.79}) \\
\enddata
\end{deluxetable}

\begin{deluxetable}{crrr}
\footnotesize \tablecaption{Same as Table \ref{tab:e_crit_1.8}
for a planet of $1M_J$, but with a different initial radius $1.2R_J$.
The eccentricities shown here are larger than those in Table
\ref{tab:e_crit_1.8} owing to the fact that a planet of a smaller
size has larger gravitational binding energy.
\label{tab:e_crit_1.2}}
\tablewidth{0pt}
\tablehead{
\colhead{$a_{\rm final}$} & \colhead{$a_{\rm init}=0.02$ AU} &
\colhead{$a_{\rm init}=0.03$ AU} &
\colhead{$a_{\rm init}=0.04$ AU}
}
\startdata
$\ge$0.04 AU & 0.157(0.26, {\bf 0.69})/0.167(0.20, {\bf 0.70}) & 0.189(1.44,
{\bf 0.84})/0.200(1.31, {\bf 0.85}) & \\
$\ge$0.05 AU & 0.168(0.36, {\bf 0.62})/0.188(0.28, {\bf 0.62}) & 0.195(1.31,
{\bf 0.75})/0.209(1.05, {\bf 0.76}) &
0.253(7.00, {\bf 0.84})/0.285(6.62, {\bf 0.81}) \\
$\ge$0.06 AU & 0.174(0.65, {\bf 0.56})/0.199(0.50, {\bf 0.57}) & 0.203(1.18,
{\bf 0.69})/0.223(0.80, {\bf 0.69}) &
0.255(6.68, {\bf 0.79})/0.286(6.17, {\bf 0.78}) \\
\enddata
\end{deluxetable}

\begin{deluxetable}{crrr}
\footnotesize \tablecaption{Same as Table \ref{tab:e_crit_1.8},
but with a different initial planet mass $0.63M_J$ and a
different initial radius $2 R_J$. The eccentricities shown here
are in general
smaller than those listed in Table \ref{tab:e_crit_1.8} and
Table \ref{tab:e_crit_1.2} except for the cases starting with
$a_{init}=0.02$ AU.
\label{tab:e_crit_63}}
\tablewidth{0pt}
\tablehead{
\colhead{$a_{\rm final}$} & \colhead{$a_{\rm init}=0.02$ AU} &
\colhead{$a_{\rm init}=0.03$ AU} &
\colhead{$a_{\rm init}=0.04$ AU}
}
\startdata
$\ge$0.04 AU & 0.124(0.58, {\bf 0.44})/0.136(0.27, {\bf 0.44}) & 0.122(0.42,
{\bf 0.54})/0.137(0.29, {\bf 0.54}) & \\
$\ge$0.05 AU & 0.145(0.87, {\bf 0.39})/0.177(0.33, {\bf 0.39}) & 0.135(0.50,
{\bf 0.49})/0.160(0.33, {\bf 0.48}) &
0.150(1.18, {\bf 0.55})/0.173(0.71, {\bf 0.56}) \\
$\ge$0.06 AU & 0.171(1.16, {\bf 0.36})/0.232(0.36, {\bf 0.36}) & 0.153(0.31,
{\bf 0.44})/0.194(0.36, {\bf 0.44}) &
0.159(1.08, {\bf 0.51})/0.190(0.60, {\bf 0.51}) \\
\enddata
\end{deluxetable}

\clearpage

\begin{figure}
\plotone{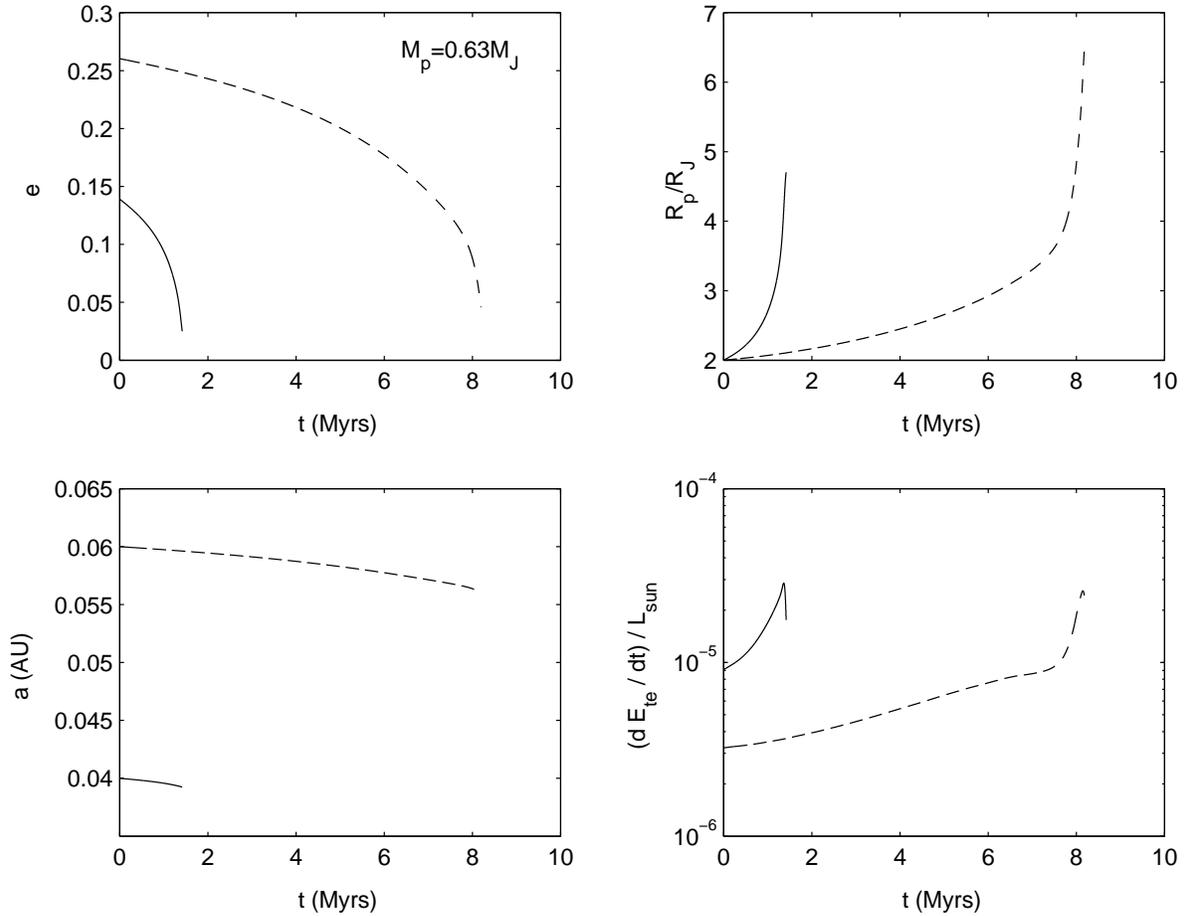}
\caption{Time evolutions of $e$, $R_p$, $a$, and $\dot E_{te}$
for a planet of a original $M_p=0.63M_J$
without a solid core initially located
at 0.04 AU (solid lines) and 0.06 AU (dashed lines).
Starting with $e\approx 0.14$ ($e=\approx 0.26$), the
planet at $a=0.04$ AU ($a=0.06$ AU) spends $t\approx 1.4$
Myrs ($t\approx 8.2$ Myrs) and then is able to
expand up to its
equilibrium radius which is just equal to its $R_L$.
\label{fig:uptoRoche}}
\end{figure}

\begin{figure}
\plotone{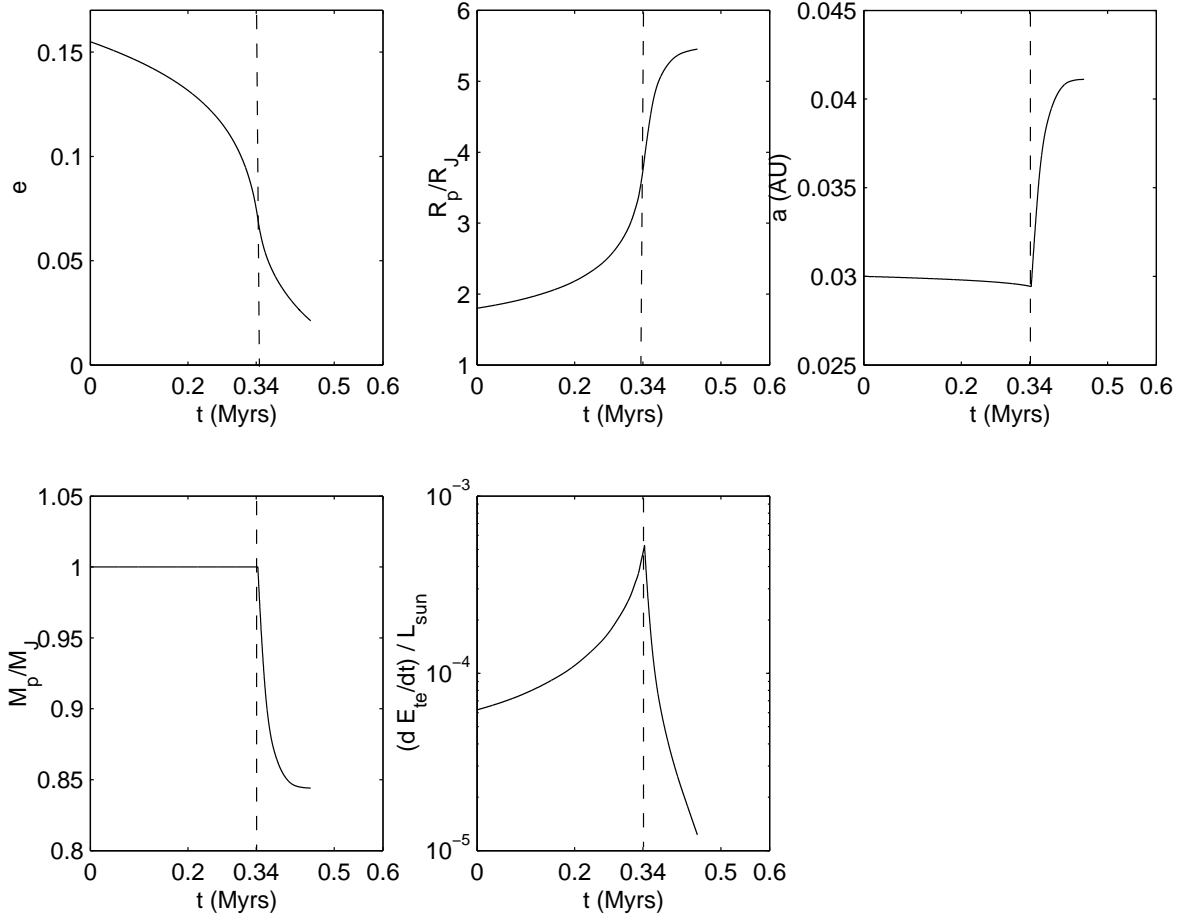}
\caption{Time evolution of $e$, $R_p$, $a$,  $M_p$, and $\dot E_{te}$
for a planet with original $M_p=1M_J$
without a solid core initially located at 0.03 AU.
The mass loss is calculated by using equation (\ref{eq:dotr2c})
with the condition that $\lambda =2$. The moment $t=0.34$ Myrs
marked by a vertical dashed line indicates the beginning of
the Roche overflow phase.
\label{overflow}}
\end{figure}

\begin{figure}
\plotone{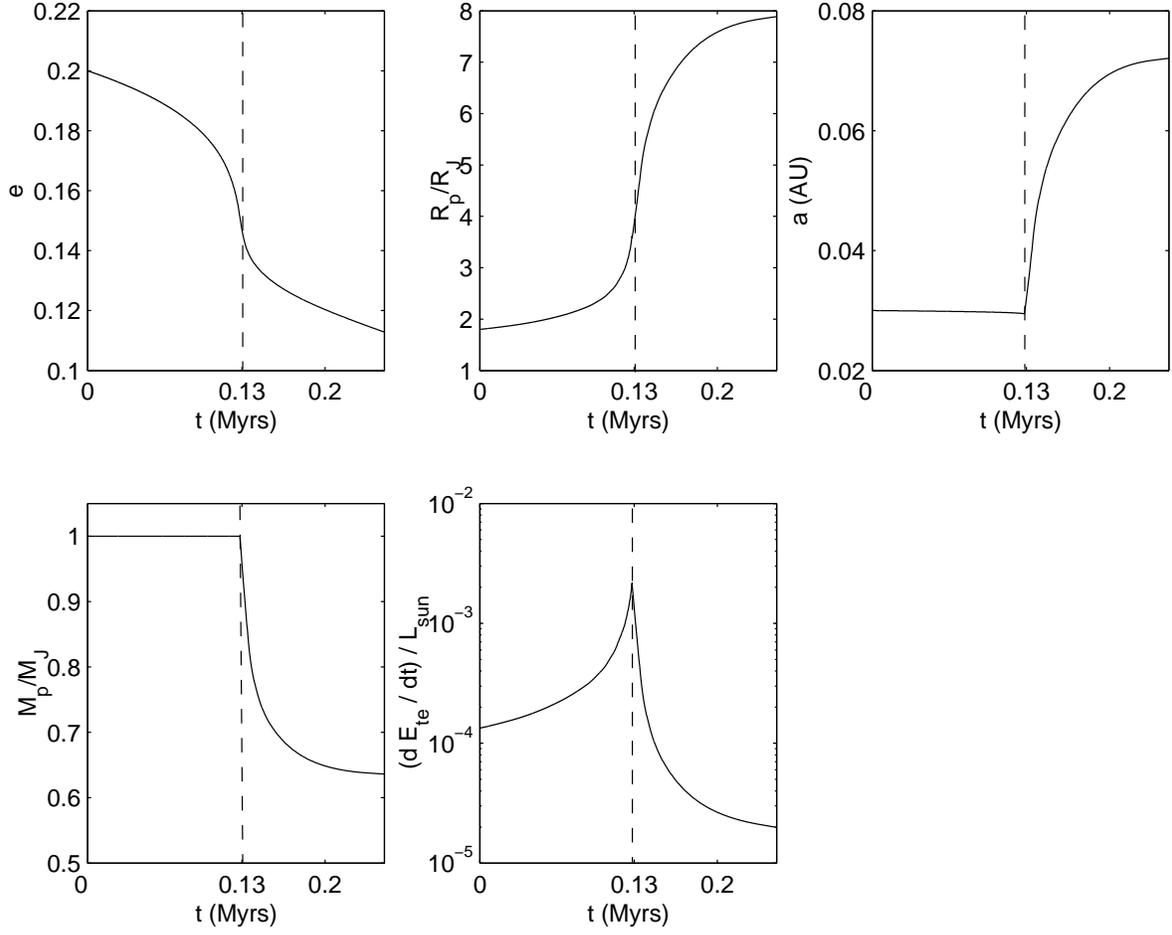}
\caption{The same simulation as the one shown in Fig. \ref{overflow}
except starting with a higher value of $e=0.2$. The moment $t=0.13$ Myrs
marked by a vertical dashed line indicates the beginning of
the Roche overflow phase in this case.
\label{overflow2}}
\end{figure}

\begin{figure}
\plotone{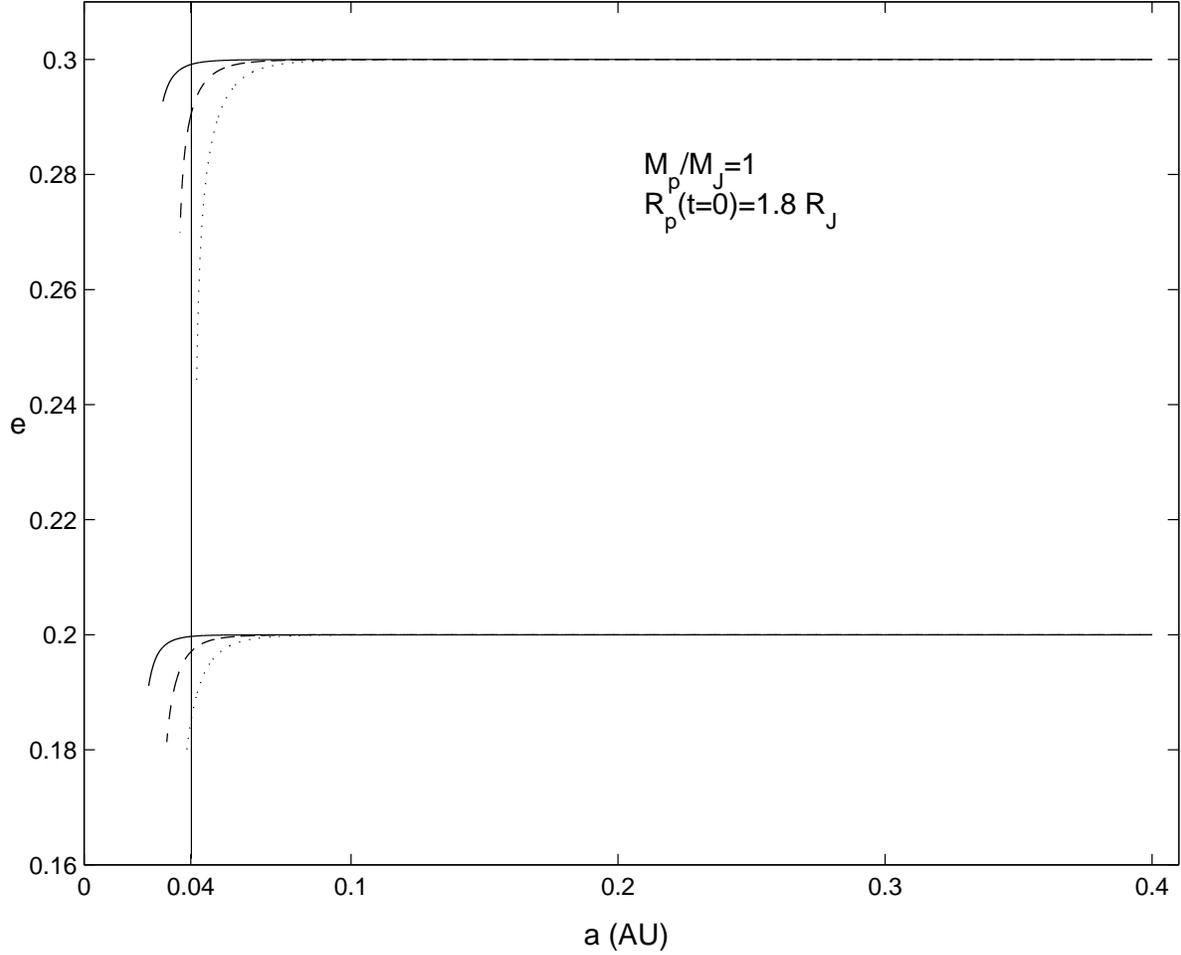}
\caption{Evolution of $a$ for a planet of 1 M$_J$ and
1.8 R$_J$ without a core and starting at $a = 0.4$ AU, 
 moving inwards at three
migration speeds: $-a/\dot a=$ 5 Myr (dotted lines), 
1 Myr (dashed lines), and 0.1 Myr (solid lines). 
The simulations start with
$e=0.2$ and $e=0.3$ and end before planets reach their
Roche lobes.
\label{edecay}}
\end{figure}

\begin{figure}
\plotone{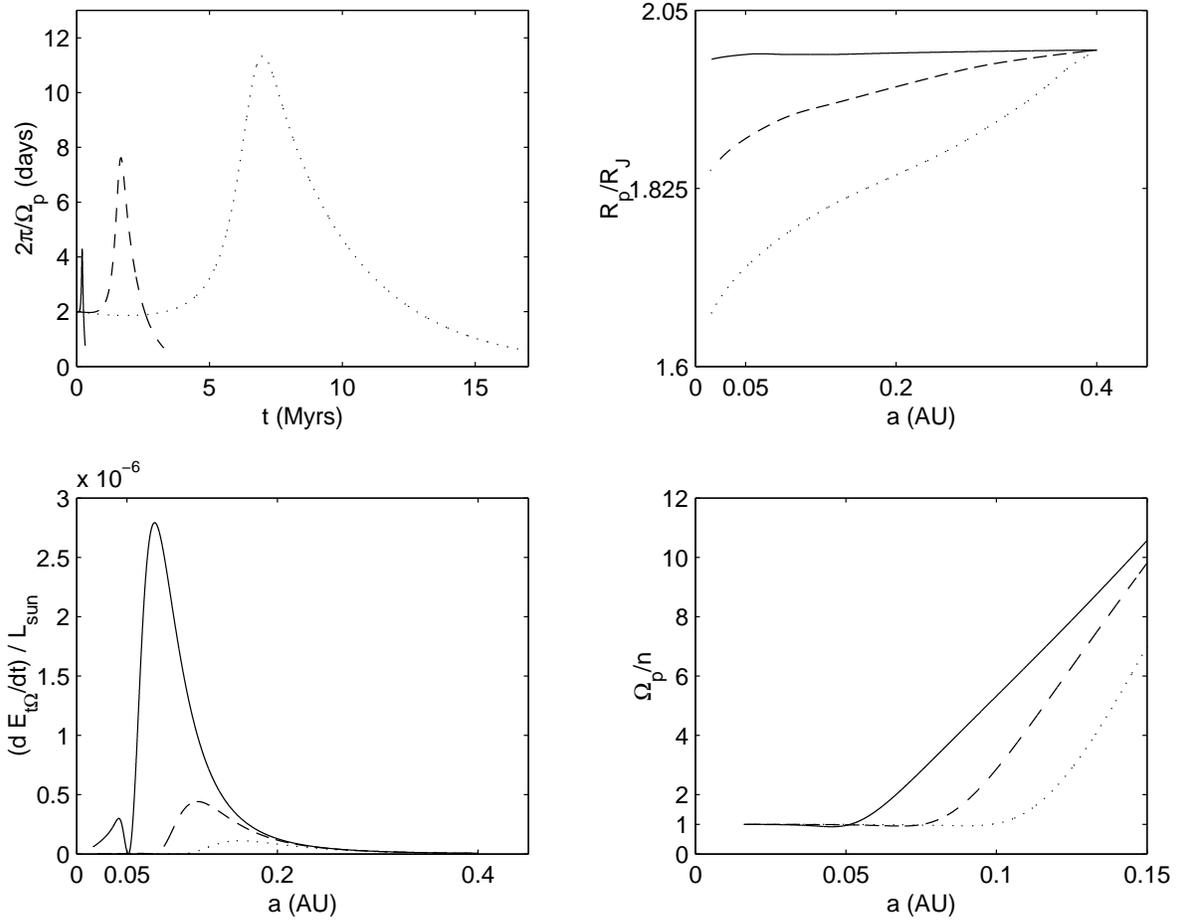}
\caption{Time evolution of $\Omega_p$, $R_p$, $\dot E_{t\Omega}$,
and $\Omega_p /n$ for a 0.63 $M_J$ planet with no core 
starting from 0.4 AU with three different migration
rates: $-a/\dot a=$5 Myr (dotted lines), 1 Myr (dashed lines), 
0.1 Myr (solid lines).
\label{sync}}
\end{figure}

\begin{figure}
\plotone{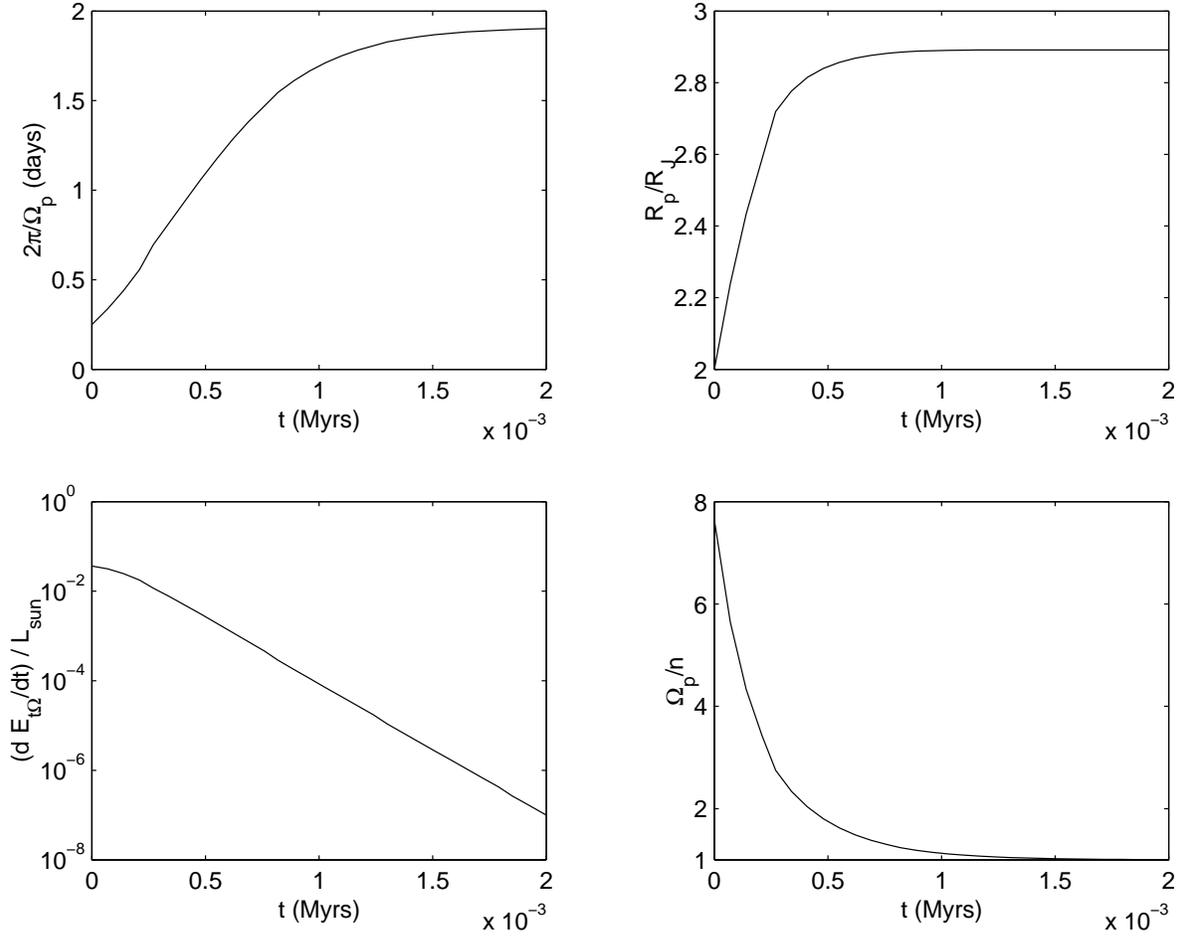}
\caption{Time evolution of $\Omega_p$, $R_p$, $\dot E_{t\Omega}$,
and $\Omega_p /n$ for a 0.63 $m_J$ planet with no core
located at 0.03 AU with no migration. The planet starts out
as a fast rotator with the ratio $\Omega_p /n \approx 7.64$,
and then is rapidly synchronized ($\Omega_p /n \rightarrow 1$)
by the tidal effect within several thousand years.
\label{sync_0.03AU}}
\end{figure}


\clearpage

\begin{thebibliography}{}
\bibitem[Applegate \& Shaham 1994]{as94} 
     Applegate J. H. \& Shaham J. 1994, \apj, 436, 312
\bibitem[Armitage et al. 2002]{allp} Armitage, P. J., Livio, M.,
     Lubow, S. H. \& Pringle J. E. 2002, \mnras, 334, 248
\bibitem[Artymowicz 1992]{art92} Artymowicz, P. 1992, \pasp, 104, 769
\bibitem[Bodenheimer, Lin, \& Mardling 2001]{peter} Bodenheimer P., Lin, D. N. C.,
    \& Mardling, R. A. 2001, \apj, 548, 466
\bibitem[Burkert \& Lin 2001]{andy} Burkert A., \& Lin, D. N. C. 2001, in
    preparation
\bibitem[Burrows et al. 1997]{bur97} Burrows, A., Marley, M., Hubbard, W. B.,
    Lunine, J. I., Guillot, T., Saumon, D., Freedman, R., Sudarsky, D.,
    \& Sharp, C. 1997, \apj, 491, 856
\bibitem[Cumming, Marcy, \& Butler 1999]{cmb01} Cumming, A., Marcy, G.W.,
    Butler, R.P. 1999, \apj, 526, 890
\bibitem[Davies \& Sigurdsson 2001]{ds01} Davies M. B., \& Sigurdsson, S.
    2001, \mnras, 324, 612
\bibitem[Dobbs-Dixon, Lin, \& Mardling 2002]{dlm02} Dobbs-Dixon, I.,
    Lin, D. N. C., \& Mardling, R. A. 2002, in preparation
\bibitem[Eggleton et al. 1998]{egg98} Eggleton, P. P., Kiseleva, L. G.,
    \& Hut, P. 1998, \apj, 499, 853
\bibitem[Ford, Rasio, \& Sills 1999]{ford99} Ford, E.B., Rasio, F.A., \& 
    Sills, A. 1999, \apj, 514, 411
\bibitem[Gilliland et al. 2001]{47Tuc} Gilliland, R. L. et al.
    2000, \apjl, 545L, 47
\bibitem[Goldreich \& Nicholson 1977]{gn77} Goldreich, P. \&
    Nicholson, P. D. 1977, Icarus, 30, 301
\bibitem[Goldreich \& Nicholson 1989]{gn89} Goldreich, P. \&
    Nicholson, P. D. 1989, \apj, 342, 1079
\bibitem[Goldreich \& Tremaine 1980]{gt80} Goldreich, P. \&
    Tremaine, S. 1980, \apj, 241, 425
\bibitem[Goldreich \& Sari 2002]{gs02} Goldreich, P. \&
    Sari, R. 2002, submitted to \apj
\bibitem[Gonzales \& Laws 2000]{gonzales00} Gonzales, 
    G., \& Laws, C. 2000, \aj, 119, 390
\bibitem[Goldreich \& Soter 1966]{gs66} Goldreich, P. \&
    Soter, S. 1966, Icarus, 5, 375
\bibitem[Goodman \& Oh 1997]{go97} Goodman, J. \& Oh, S. 1997,
     \apj, 486, 403
\bibitem[Hubbard 1984]{hub84} Hubbard, W. B. (1984). Planetary interiors.
      New York, Van Nostrand Reinhold Co.
\bibitem[Ioannou \& Lindzen 1993a]{il93a} Ioannou, P. J. \& Lindzen, R. S.
     1993, \apj, 406, 252
\bibitem[Ioannou \& Lindzen 1993b]{il93b} Ioannou, P. J. \& Lindzen, R. S.
     1993, \apj, 406, 266
\bibitem[Ioannou \& Lindzen 1994]{il94} Ioannou, P. J. \& Lindzen, R. S.
     1994, \apj, 424, 1005
\bibitem[Israelian et al. 2001]{Li6} Israelian, G., Santos, N. C.,
    Mayor, M. \& Rebolo, R. 2001, Nature, 411, 163
\bibitem[Jiang et al. 2003]{jiang03} Jiang, I.-G., Ip, W.-H., \& Yeh, L.-C.,
     ApJ, 582, 449, 2003
\bibitem[Johns-Krull et al. 1999]{jk99} Johns-Krull, C. M.,
    Valenti, J. A., \& Koresko, C. 1999, \apj, 516, 900
\bibitem[Konigl 1991]{kon91} Konigl, A. 1991, \apjl, 370L, 39
\bibitem[Kuchner \& Lecar 2002]{kl02} Kuchner, M. J. \& Lecar, M. 2002,
    to appear in ApJL
\bibitem[Lin 1997]{lin97} Lin, D,N.C. 1997,
        in Accretion Phenomena and Related Outflows; IAU Colloquium 163. ASP
        Conference Series; Vol. 121; 1997; ed. D. T.
	        Wickramasinghe; G. V. Bicknell; and L. Ferrario (1997), p.321		
\bibitem[Lin et al. 1996]{lin96} Lin, D. N. C., Bodenheimer, P.,
    \& Richardson, D. C. 1996, Nature, 380, 606
\bibitem[Lin \& Ida 1997]{li97} Lin, D. N. C. \& Ida, S. 1997,
     \apj, 477, 781
\bibitem[Lin \& Papaloizou 1986]{lp86} Lin, D. N. C. \&
    Papaloizou, J. C. B. 1986, \apj, 309, 846
\bibitem[Lubow \& Shu 1975]{ls75} Lubow, S. H. \& Shu, F. 1975, 
     \apj, 198, 383
\bibitem[Lubow et al.(1997)]{lub97} Lubow, S. H., Tout, C. A.,
      \& Livio, M. 1997, \apj, 484, 866
\bibitem[Marcy et al. 2000]{marcy00} Marcy, G. W., Cochran, W. D.,
    \& Mayor, M. 2000, in Protostars and Planets IV, ed. V. Mannings,
    A. P. Boss, \& S. Russell (Tucson: Univ. Arizona Press), 1285
\bibitem[Mardling \& Lin 2002]{ml02} Mardling, R. A. \&
     Lin, D. N. C., ApJ, 573, 829, 2002
\bibitem[Mathieu et al. 1992]{mat92} Mathieu, R. D., Duquennoy, A., 
     Latham, D. W., Mayor, M., Mazeh, T., \& Mermilliod, J.-C. 1992, 
     in Binaries as Tracers of Stellar Formation, ed. A. Duquennoy 
     \& M. Mayor (Cambridge: Cambridge Univ. Press), 278
\bibitem[Mathieu 1994]{mat94} Mathieu, R. D. 1994, \araa, 32, 465
\bibitem[Mayor \& Queloz 1995]{mq95} Mayor, M. \& Queloz, D. 1995,
     Nature, 378, 355
\bibitem[Murray \& Dermott 1999]{md99} Murray C. D. and
    Dermott S. F. 1999, {\it Solar System Dynamics}, Cambridge
    Univ. Press
\bibitem[Nagasawa et al. 2002]{na02} Nagasawa, M., Lin, D. N. C.,
    \& Ida, S. 2002, submitted to \apj
\bibitem[Oh \& Lin 2002]{kapsoo} Oh, K. \& Lin, D. N. C. 2002, in preparation
\bibitem[Paetzold \& Rauer 2002]{pr02} Paetzold, M. \& Rauer, H. 2002, 
    \apjl, 568, 117
\bibitem[Papaloizou et al. 2001]{pap01} Papaloizou, J. C. B.,
    Nelson, R. P., \& Masset, F. 2001, \aap, 366, 263
\bibitem[Peale et al. 1979]{peale} Peale, S. J., Cassen, P.
    \& Reynolds, R. T. 1979, Science, 203, 892
\bibitem[Pollack et al. 1996]{pol96} Pollack, J. B., Hubickyj, O.,
     Bodenheimer, P., Lissauer, J. J., Podolak, M., \& Greenzweig, Y.
     1996, Icarus, 124, 62
\bibitem[Pringle 1985]{pri85} Pringle, J.E. 1985, in {\it Interacting Binary
     Stars}, eds J.E. Pringle \& R.A. Wade, Cambridge University Press:
     Cambridge
\bibitem[Rasio et al. 1996]{ra96} Rasio, F. A., Tout, C. A., Lubow, S. H,
    \& Livio M, ApJ, 470, 118, 1996
\bibitem[Rasio \& Ford 1996]{rf96}  Rasio, F. A. \& Ford, E. B. 1996,
     Science, 274, 954
\bibitem[Sandquist et al. 2002]{san02} Sandquist, E. L., Dokter, J. J.,
     Lin, D. N. C., \& Mardling, R. A. 2002, \apj, 572, 1012
\bibitem[Sandquist et al. 1998]{san98} Sandquist, E. L., Taam, R. E.,
     Lin, D. N. C., \& Burkert, A. 1998, \apjl, 506L, 65
\bibitem[Showman \& Guillot 2002]{sg02} Showman, A. P. \& Guillot, T.
     2002, \aap, 385, 166
\bibitem[Shu et al. 1994]{shu94} Shu, F., Najita, J., Ostriker, E.,
     Wilkin, F., Ruden, S., \& Lizano, S. 1994, \apj, 429, 781
\bibitem[Stassun et al. 1999]{sta99} Stassun, K. G., Mathieu, R. D.,
     Mazeh, T., \& Vrba, F. J. 1999, \aj, 117, 2941
\bibitem[Takeuchi et al. 1996]{taku} Takeuchi, T., Miyama, S. M.,
    \& Lin, D. N. C. 1996, \apj, 460, 832
\bibitem[Terquem et al. 1998]{ter98} Terquem, C., Papaloizou, J. C. B.,
    Nelson, R. P., \& Lin, D. N. C. 1998, \apj, 502, 788
\bibitem[Trilling et al. (1998)]{tri98} Trilling, D. E.,
    Benz, W., Guillot, T., Lunine, J. I., Hubbard, W. B.,
    \& Burrows, A. 1998, \apj, 500, 428
\bibitem[Ward 1981]{ward81} Ward, W. R. 1981, Icarus, 47, 234
\bibitem[Weidenschilling \& Marzari 1996]{wm96} Weidenschilling, S. J.
    \& Marzari, F. 1996, Nature, 384, 619
\bibitem[Wuchterl, Guillot, \& Lissauer 2000]{wuch00} 
    Wuchterl, G., Guillot, T., \& Lissauer, J.J. 2000, in Protostars and
    Planets IV, ed V. Mannings, A. P. Boss, \& S. S. Russell (Tucson:
    Univ. of Arizona Press), 1081
\bibitem[Yoder \& Peale 1981]{yp81} Yoder, C. F. \& Peale, S. J.
    1981, Icarus, 47, 1
\bibitem[Zahn 1977]{zahn77} Zahn, J.-P. 1977, \aap, 57, 383
\bibitem[Zahn 1989]{zahn89} Zahn, J.-P. 1989, \aap, 220, 112
\bibitem[Zucker \& Mazeh 2002]{zm02} Zucker, S. \& Mazeh, T.
    2002, \apjl, 568L, 113
\end{thebibliography}
\end{document}